\documentclass[twoside]{dis04}
\def\runauthor{Andr\'e Sopczak}
\def\shorttitle{Searches for Higgs Bosons Beyond the Standard Model}

\newcommand{\ee}{\mbox{$\mathrm{e}^{+}\mathrm{e}^{-}$}}
\newcommand{\ra}{\mbox{$\rightarrow$}}

\newcommand{\Zo} {{\mathrm {Z}}}
\newcommand{\db}    {{d_{\rm B}}}%
\newcommand{\dgz}  {{\Delta g_1^{\Zo}}}%
\newcommand{\dkg}   {{\Delta \kappa_\gamma}}%

\begin{document}

\thispagestyle{empty}
%\begin{titlepage}
\def\thefootnote{\fnsymbol{footnote}}       % symbols for footnotes

\begin{center}
\mbox{ }

\end{center}
\begin{flushright}
\Large
\mbox{\hspace{10.2cm} hep-ph/0408047} \\
%%\mbox{\hspace{12.0cm} August 2004}
\end{flushright}
\begin{center}
\vskip 1.0cm
{\Huge\bf
Searches for Higgs Bosons \\
       Beyond the Standard Model
}
\vskip 1cm
{\LARGE\bf Andr\'e Sopczak}\\
\smallskip
\Large Lancaster University

\vskip 2.5cm
\centerline{\Large \bf Abstract}
\end{center}

\vskip 2.cm
\hspace*{-1cm}
\begin{picture}(0.001,0.001)(0,0)
\put(,0){
\begin{minipage}{\textwidth}
\Large
\renewcommand{\baselinestretch} {1.2}
Latest results from the combined data of the four LEP experiments
ALEPH, \mbox{DELPHI}, L3 and OPAL on Higgs boson searches in extensions of the
Standard Model (SM) are presented.
\renewcommand{\baselinestretch} {1.}

\normalsize
\vspace{5.5cm}
\begin{center}
{\sl \large
Presented at the XII Workshop on Deep Inelastic Scattering, DIS'2004,\\
High Tatras, Slovakia, 14--18 April 2004.
\vspace{-6cm}
}
\end{center}
\end{minipage}
}
\end{picture}
\vfill

%\end{titlepage}

%%%%%%%%%%%%%%%%%%%%%%%%%%%%%%%%%%%%%%%%%%%%%%%%%%%%%%%%%%%%%%%%%%%%%%%%%%%

\newpage
\thispagestyle{empty}
\mbox{ }
\newpage
\setcounter{page}{1}
%%%%%%%%%%%%%%%%%%%%%%%%%%%%%%%% AS end %%%%%%%%%%%%%%%%%%%%%%%%%%%%%%%%%%%

\title{Searches for Higgs Bosons \\
       Beyond the Standard Model}

\author{Andr\'e Sopczak}

\address{Lancaster University, UK. E-mail: Andre.Sopczak@cern.ch}

\maketitle

\abstracts{
Latest results from the combined data of the four LEP experiments 
ALEPH, \mbox{DELPHI}, L3 and OPAL on Higgs boson searches in extensions of the
Standard Model (SM) are presented.
}

\vspace*{-0.1cm}
The LEP experiments took data between August 1989 and November 2000
at center-of-mass energies first around the Z resonance (LEP-1) and later
up to 209 GeV (LEP-2). In 2000 most data was taken around 206 GeV.
The LEP accelerator operated very successfully and 
a total luminosity of ${\cal L} = 2461$ pb$^{-1}$ was accumulated
at LEP-2 energies.
Data-taking ended on 3 November 2000, although some data excess was 
observed in searches for the SM Higgs boson with 115~GeV mass.
In this report several different research lines are addressed
beyond the SM:
1) coupling limits;
2) the Minimal Supersymmetric extension of the SM (MSSM):
dedicated searches, three-neutral-Higgs-boson hypothesis,
benchmark and general scan mass limits;
3) CP-violating models;
4) invisible Higgs boson decays;
5) neutral Higgs bosons in the general two-doublet Higgs model;
6) Yukawa Higgs boson processes $\rm b\bar b h$ and $\rm b\bar b A$;
7) singly-charged Higgs bosons;
8) doubly-charged Higgs bosons;
9) fermiophobic Higgs boson decays $\rm h\ra WW, ZZ, \gamma\gamma$;
10) uniform and stealthy Higgs boson scenarios.

While the results from Standard Model Higgs boson searches are final~\cite{sm},
the results of searches in extended models are mostly preliminary~\cite{summer2003}.
Limits are given at 95\% CL. 
Full-size figures of this report are available from Ref.~\cite{dis04hepph}.

\vspace*{-0.3cm}
\section{Coupling Limits}
\vspace*{-0.2cm}

\begin{figure}[htb]
\vspace*{-0.85cm}
\begin{center}
\includegraphics[scale=0.21]{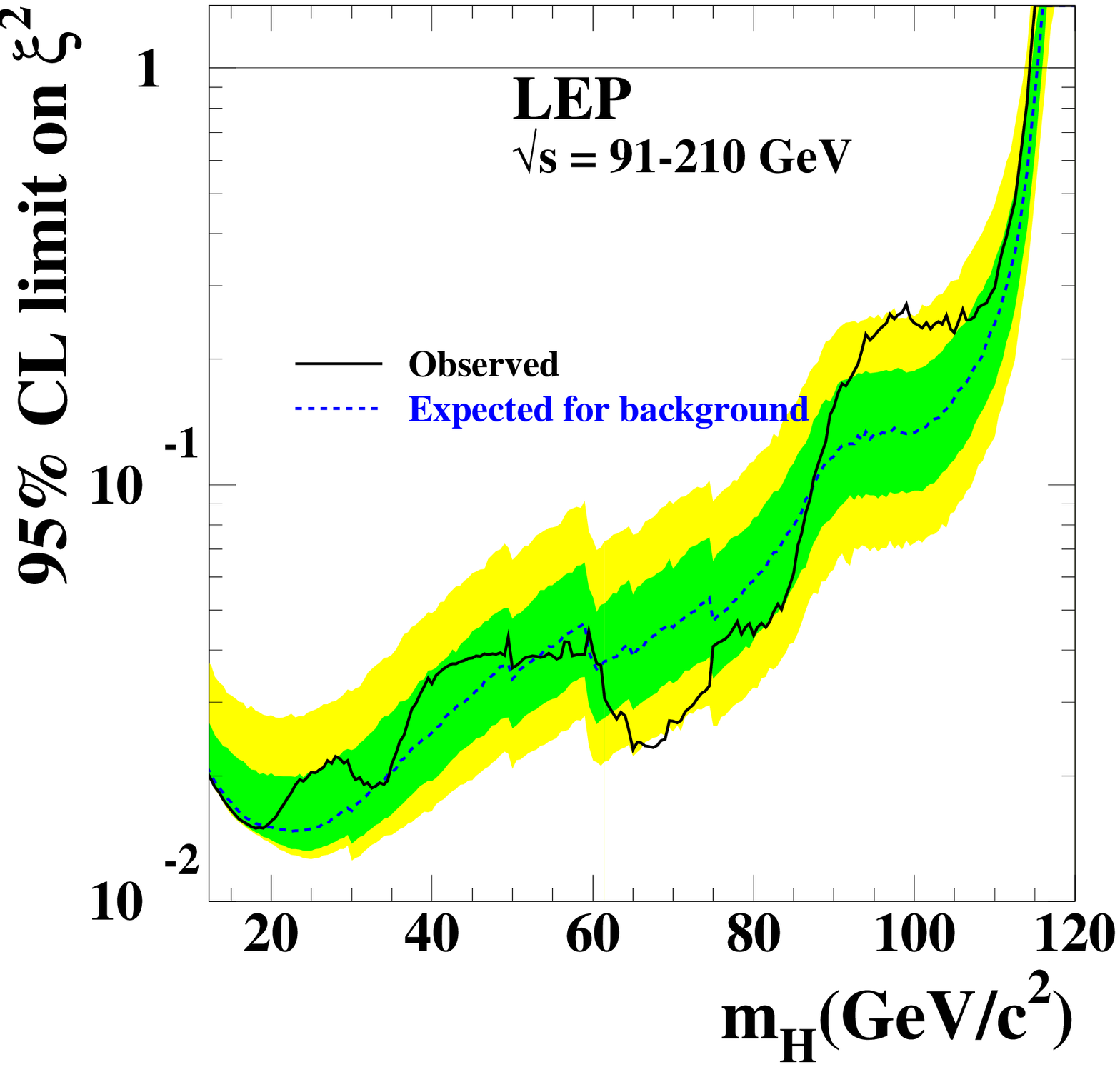}\hfill
\includegraphics[scale=0.21]{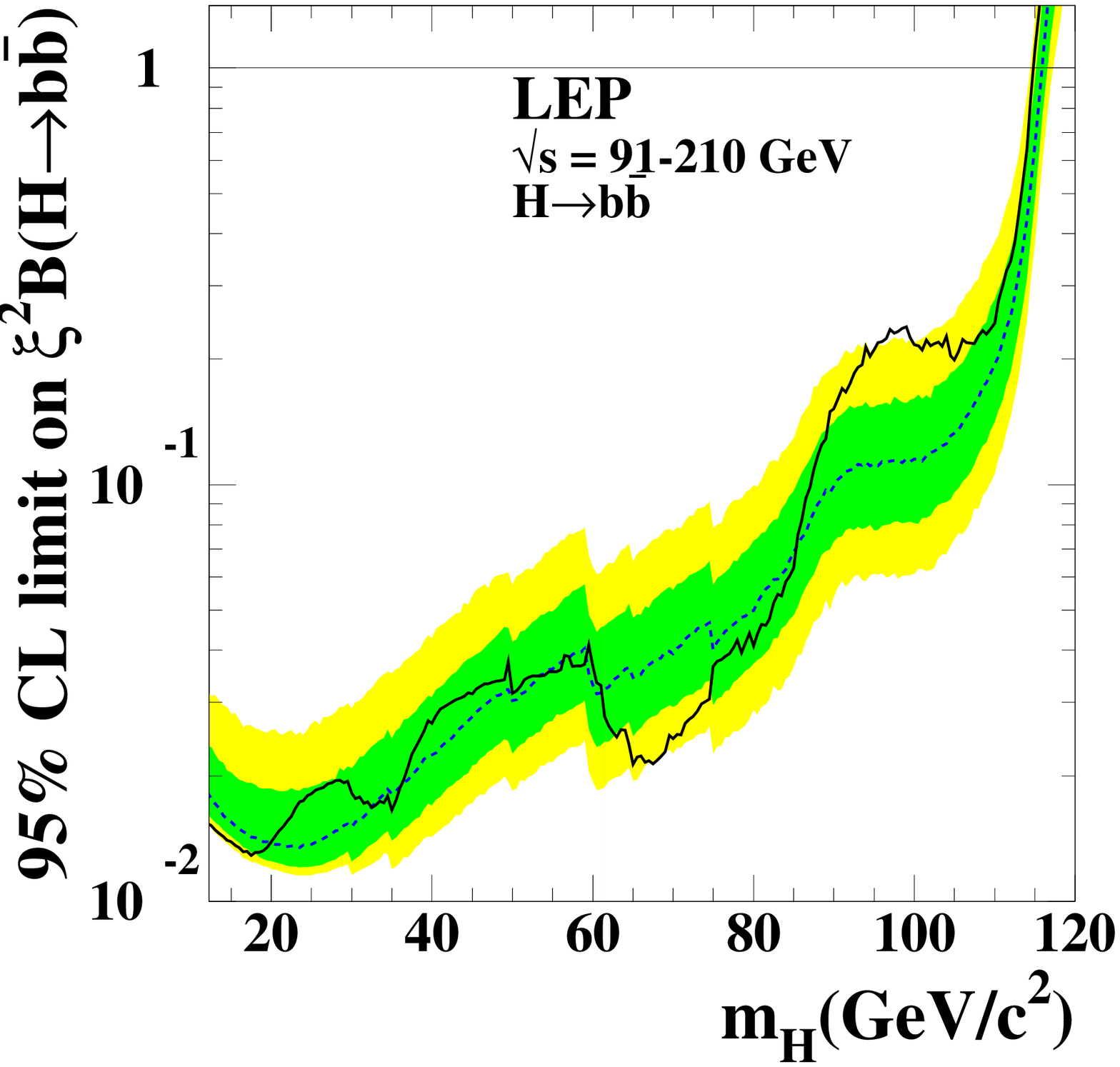}\hfill
\includegraphics[scale=0.21]{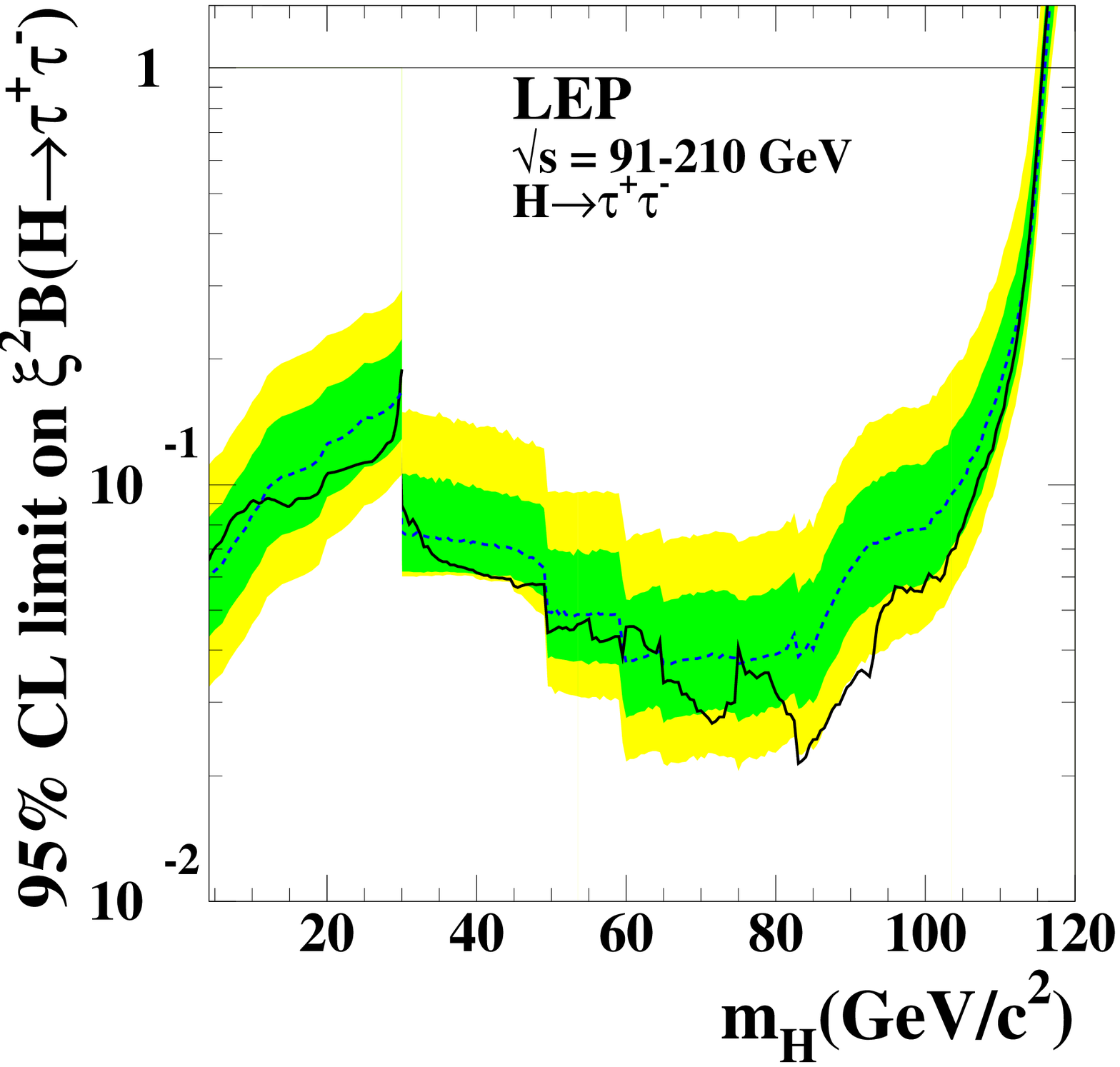}
\end{center}
\vspace*{-0.4cm}
\caption[]{Coupling limits.
Left:\,SM\,decay\,mode.\,Center:\,b-quark\,decay\,mode.\,Right:\,$\tau$-lepton\,decay mode. 
The\,1$\sigma$\,and\,2$\sigma$\,error\,bands\,on\,the\,expected\,limit for background are indicated (shaded areas).
\label{fig:limits}}
\vspace*{-0.65cm}
\end{figure}

Figure~\ref{fig:limits} shows coupling limits assuming the Higgs boson decays with 
SM branching fractions and a SM production rate reduced by 
$\xi^2 = (g_{\rm HZZ}/g_{\rm HZZ}^{\rm SM})^2$. In addition, coupling limits are presented for
b-quark and  $\tau$-lepton decay modes.
Remarkably, mass limits for flavor-blind hadronic decays are close to the SM decay mode limits.

\section{Minimal Supersymmetric Extension of the SM (MSSM)}

\vspace*{-0.2cm}
\subsection{Benchmark Limits and Dedicated Low $m_{\rm A}$ and $\rm h\ra AA$ Searches}
\vspace*{-0.2cm}

\begin{figure}[htb]
\vspace*{-0.7cm}
\begin{center}
\includegraphics[scale=0.18]{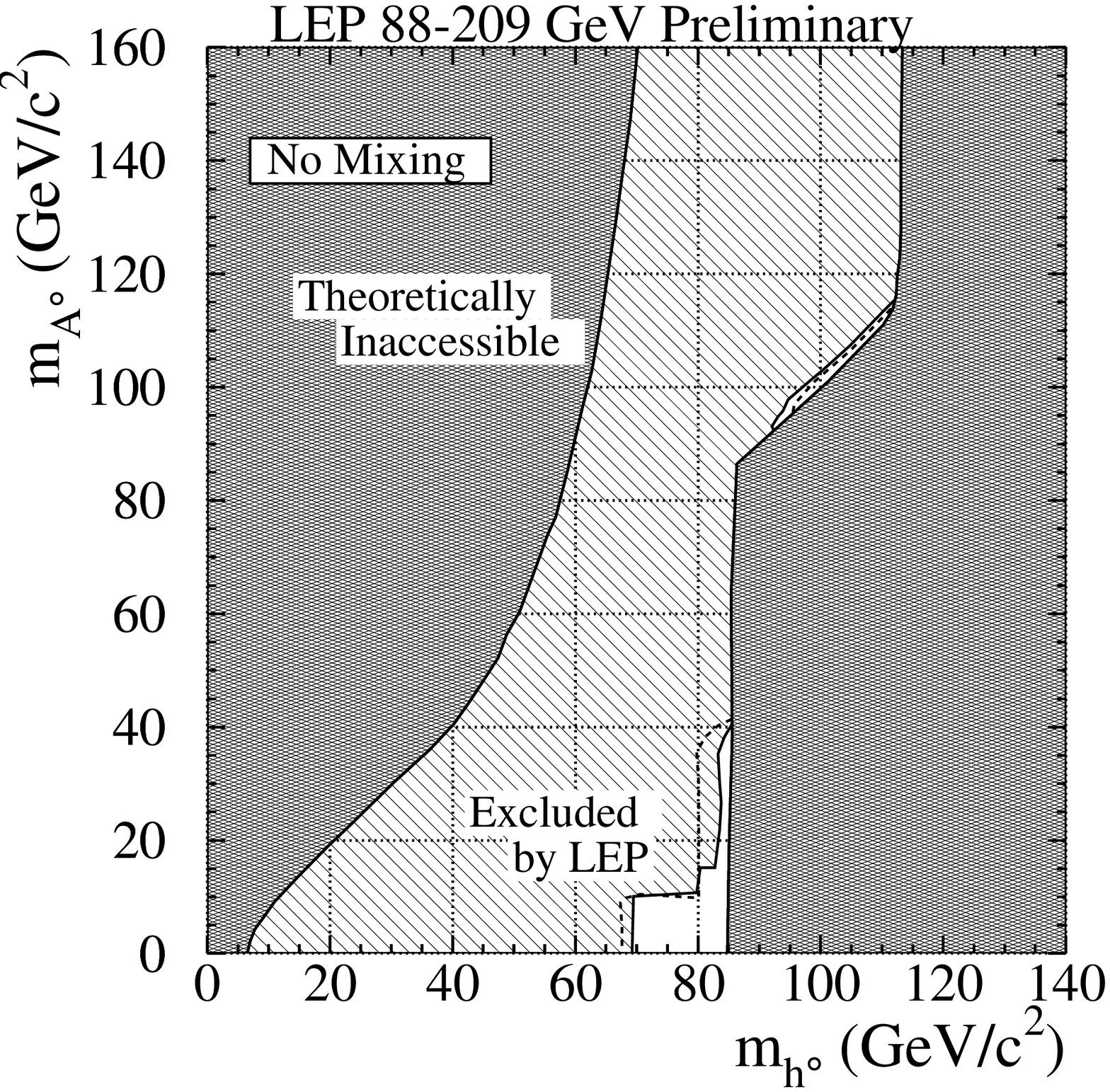}\hfill
\includegraphics[scale=0.18]{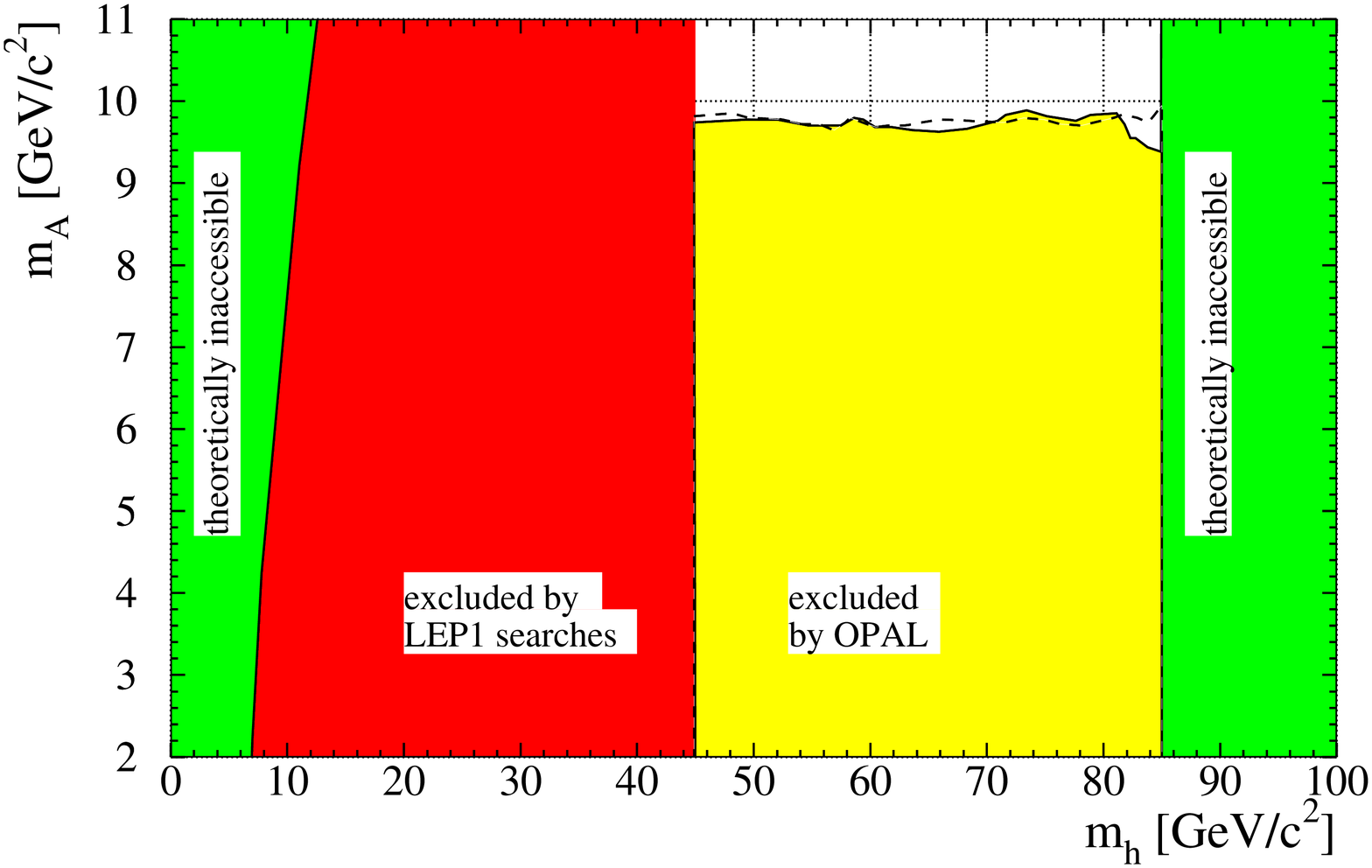}\hfill
\includegraphics[scale=0.23]{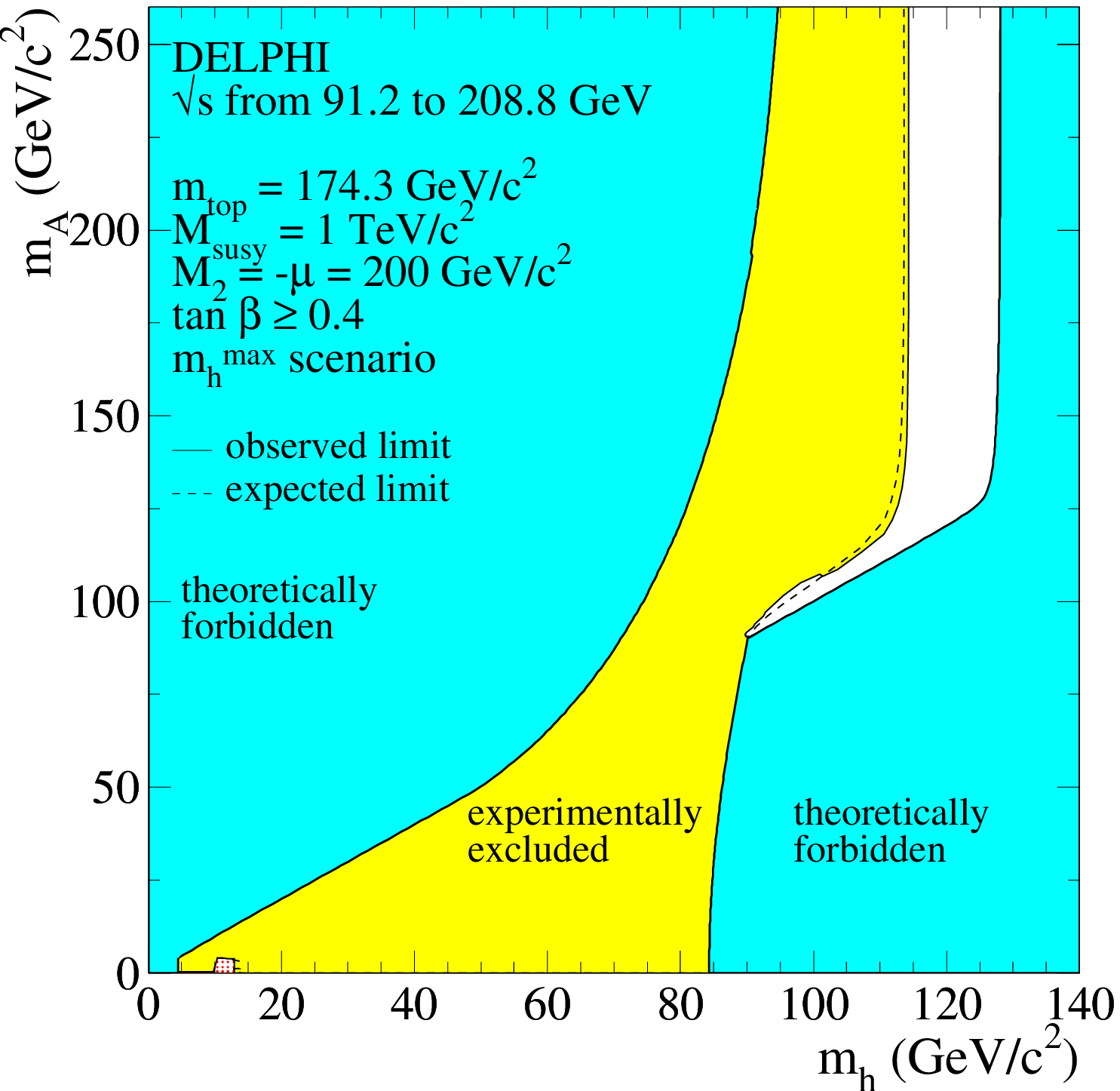}
\end{center}
\vspace*{-0.4cm}
\caption[]{MSSM. Left: unexcluded mass region for a light A boson
           in the no-mixing scalar top benchmark scenario.
           Center: excluded mass region by dedicated searches
           for a light A boson.
           Right: mass limits in the maximum h-mass benchmark scenario.                
\label{fig:light}}
\vspace*{-0.6cm}
\end{figure}

Figure~\ref{fig:light} shows a small previously unexcluded mass region
for light A masses in the no-mixing scalar top benchmark scenario.
This region is mostly excluded by new dedicated searches for 
a light A boson (center plot). Limits for the maximum h-mass benchmark 
scenario, including results from dedicated searches for the 
reaction $\rm h\ra AA$ are also shown (right plot).

\vspace*{-0.4cm}
\subsection{Three-Neutral-Higgs-Boson Hypothesis and a MSSM Parameter Scan}
\begin{figure}[htb]
\vspace*{-1.2cm}
\begin{center}
\includegraphics[scale=0.19]{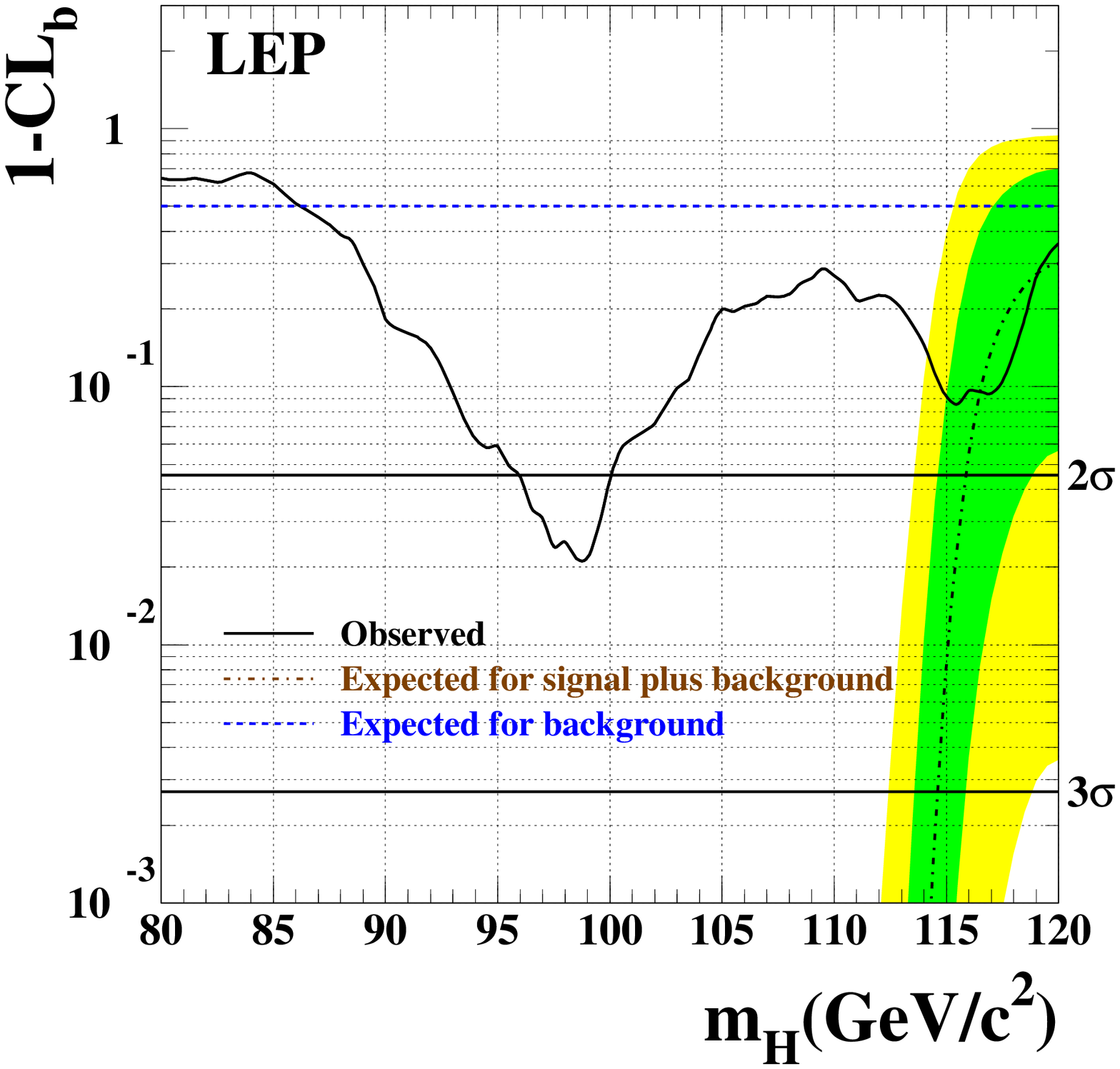}       \hfill
\includegraphics[scale=0.28]{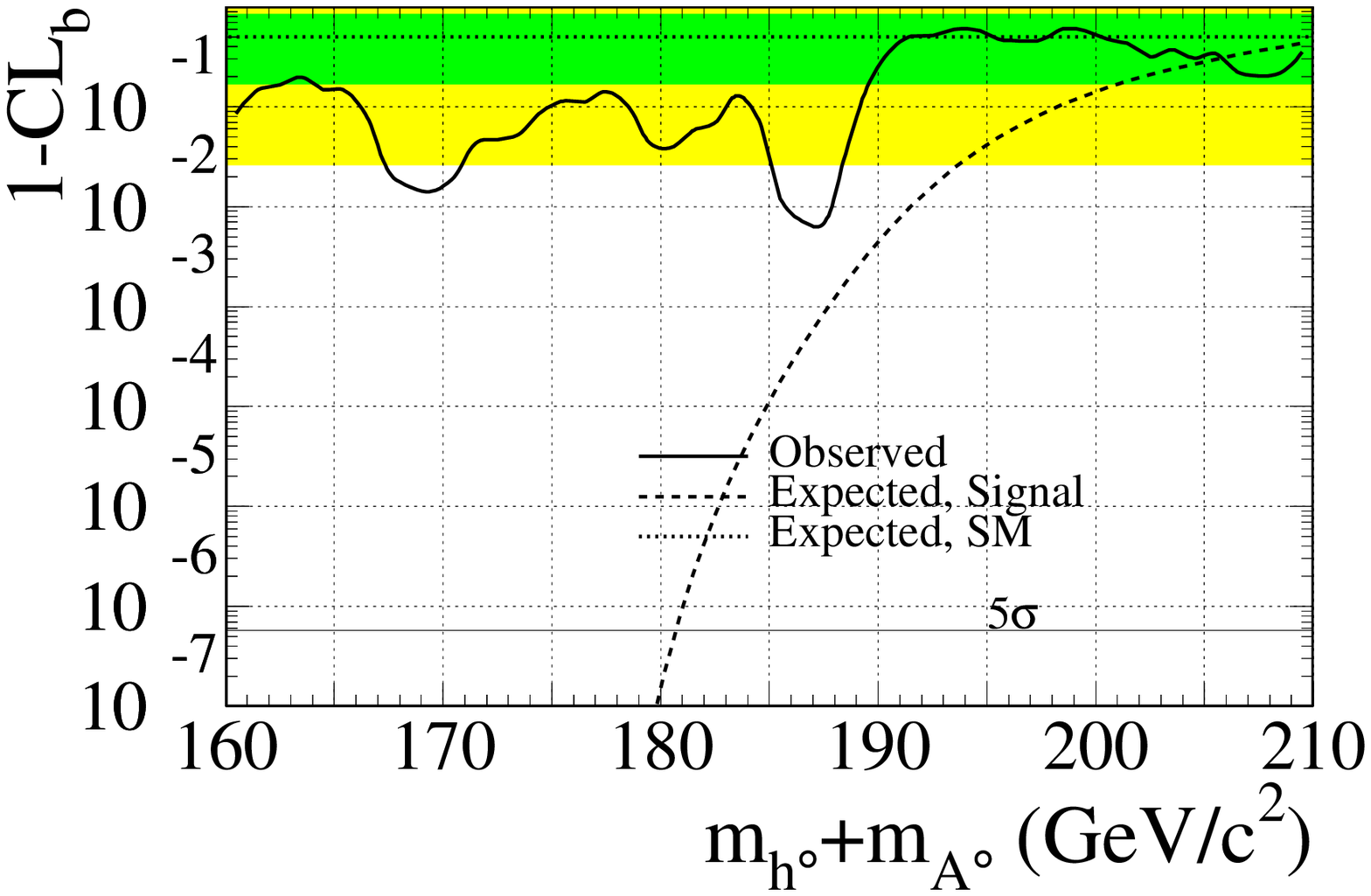} \hfill
\includegraphics[scale=0.19]{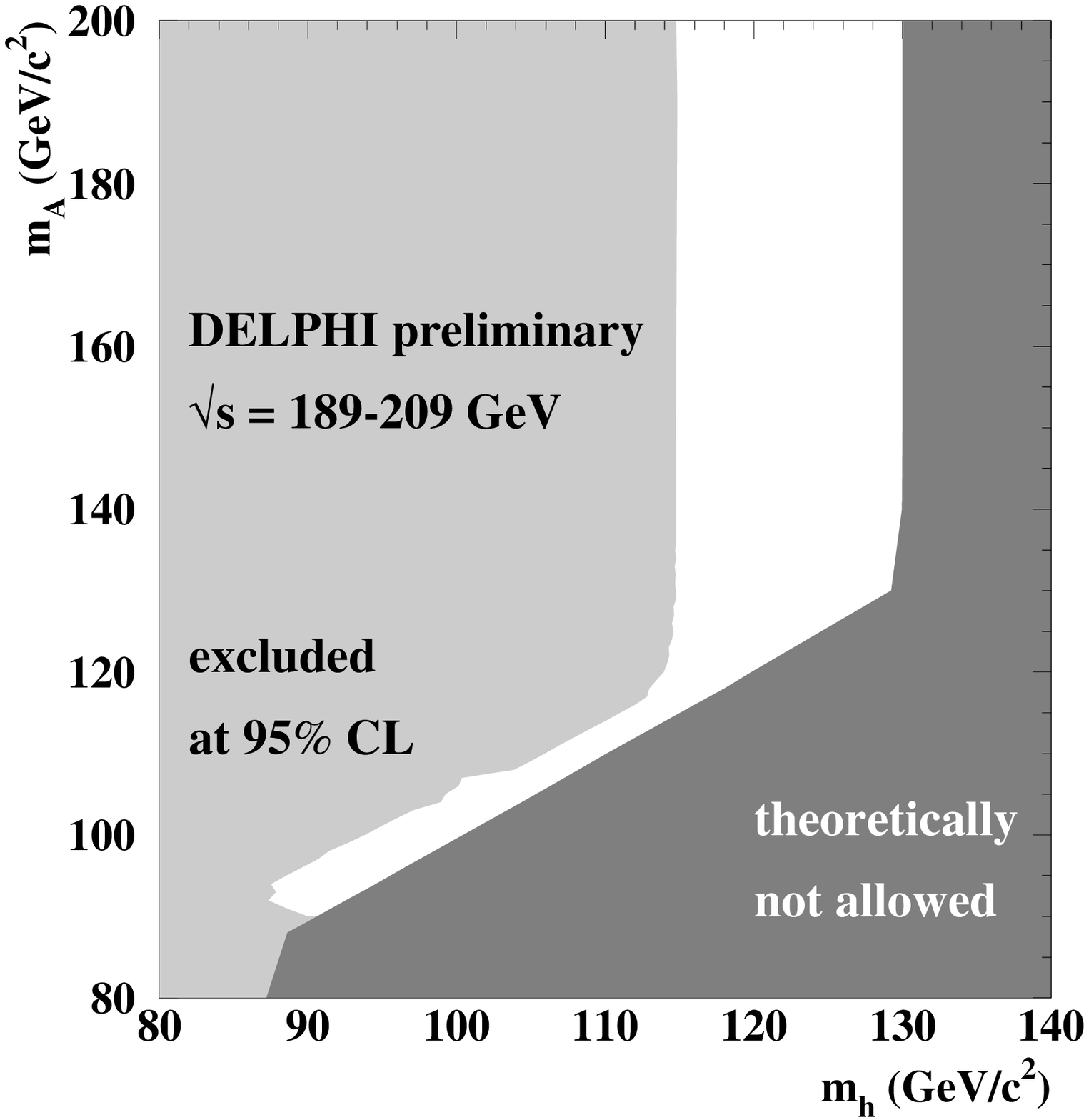}
\end{center}
\vspace*{-0.4cm}
\caption[]{MSSM. Left: small data excess at 99 GeV and 116 GeV in hZ/HZ searches.
                 Center: small data excess at 
                 $m_{\rm h}+m_{\rm A}=187$~GeV in hA searches.
$1-CL_{\rm b}$ expresses the incompatibility of the observation
with the background-only hypothesis.
                 Right: mass limits from a general MSSM parameter scan.
\label{fig:three}}
\vspace*{-0.5cm}
\end{figure}

\vspace*{-0.2cm}
The hypothesis of three-neutral-Higgs-boson production, via hZ, HZ and hA
is compatible with the data excess seen in Fig.~\ref{fig:three}.
For the reported MSSM parameters~\cite{as2000} reduced hZ production 
near 100 GeV and HZ production near 115 GeV is compatible with the 
data (left plot). 
For $m_{\rm h} \approx m_{\rm A}$, hA production is also compatible 
with the data (center plot).
The parameters have been obtained from a general MSSM parameter scan. Mass limits
from this scan are also shown (right plot).

\vspace*{-0.4cm}
\subsection{Large Effect from Increased Top-Quark Mass}
\vspace*{-0.2cm}

The increase of the measured top quark mass by about 4 GeV has a large effect\,on the 
Higgs boson mass and reduces significantly the previous limits on $\tan\beta$.

\vspace*{-0.2cm}
\section{CP-Violating Models}
\vspace*{-0.2cm}

\begin{figure}[htb]
\vspace*{-0.3cm}
\begin{center}
\includegraphics[scale=0.3]{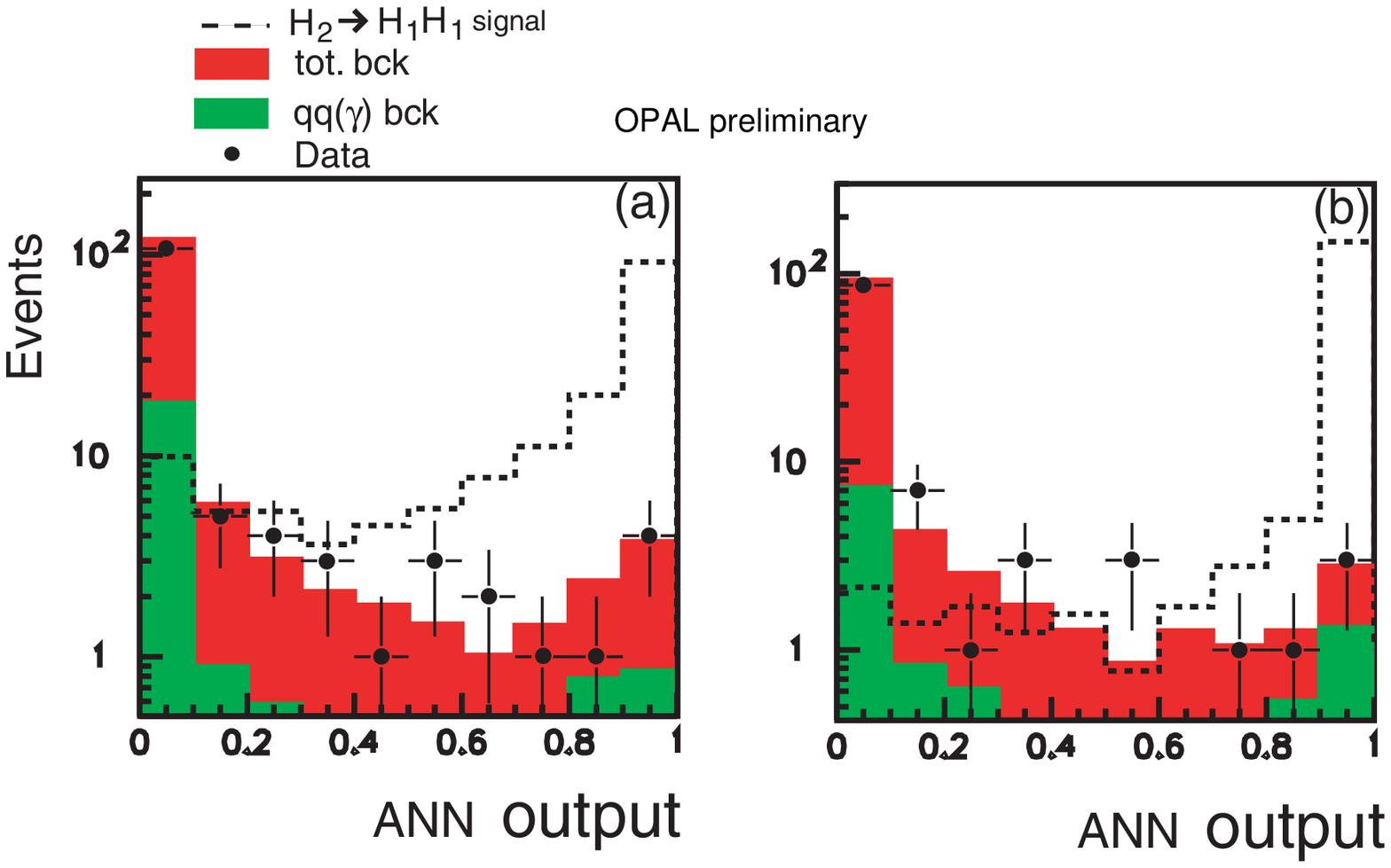}\hfill
\includegraphics[width=0.25\textwidth,bbllx=279pt,bblly=2pt,bburx=512pt,bbury=236pt,clip=]
{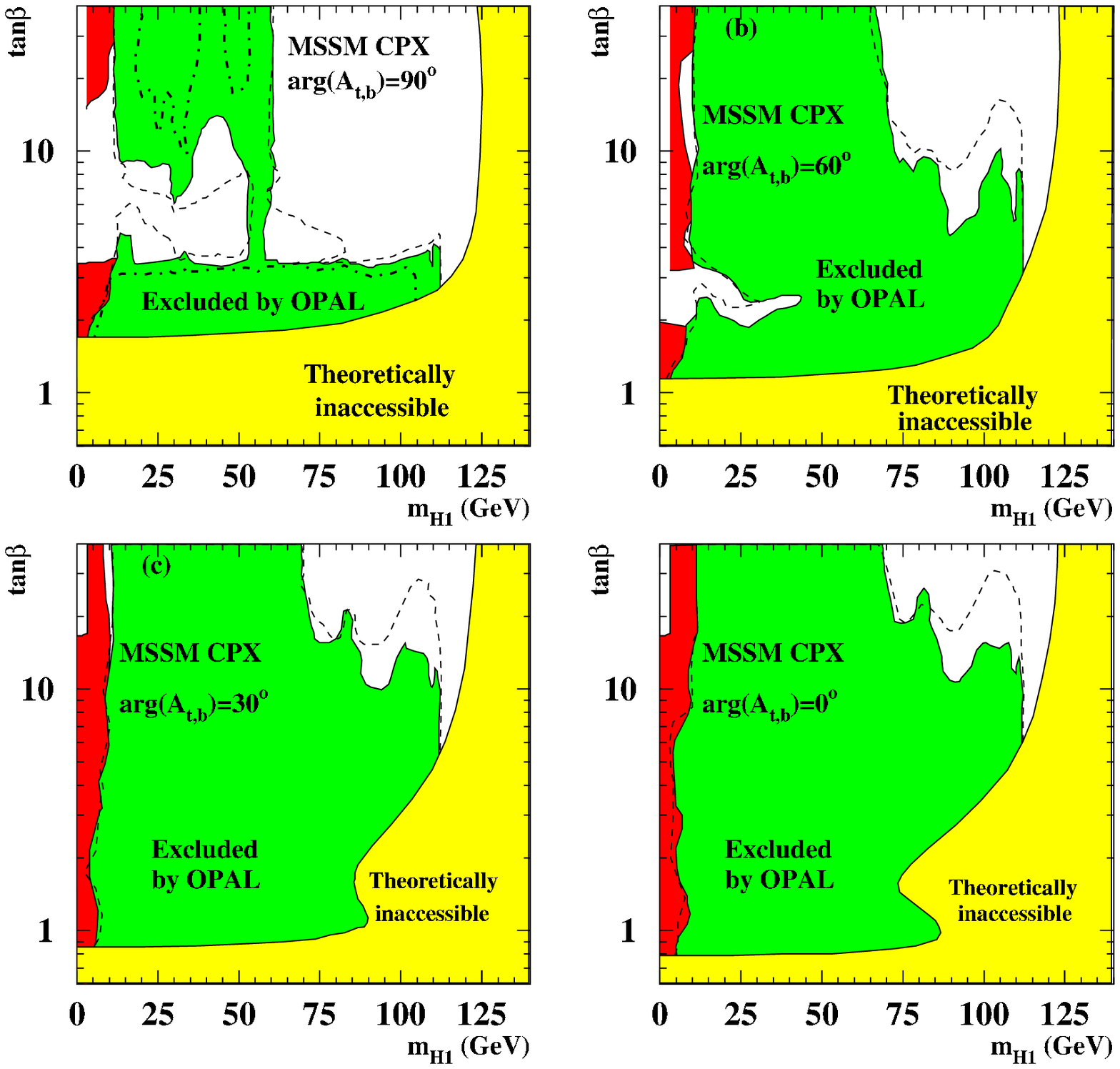}
\includegraphics[width=0.25\textwidth,bbllx=23pt,bblly=240pt,bburx=257pt,bbury=471pt,clip=]
{newo_fig25.eps}
\end{center}
\vspace*{-0.4cm}
\caption[]{CP-violation models.
Left: Artificial Neural Network (ANN) output distributions for the reactions
$\rm \ee\ra H_2 Z \ra b\bar b \nu\bar \nu$ and 
$\rm \ee\ra H_2 Z \ra H_1 H_1 Z \ra b\bar b b\bar b \nu\bar \nu$
for different data sub-samples. No indication of a signal is observed.
Center: mass limits with no CP-mixing.
Right: mass limits with full CP-mixing.
\label{fig:cpvio}}
\vspace*{-0.7cm}
\end{figure}

Instead of h, H and A, the Higgs bosons are named $\rm H_1, H_2$ and $\rm H_3$.
The reactions
$\rm \ee\ra H_2 Z \ra b\bar b \nu\bar \nu$ and 
$\rm \ee\ra H_2 Z \ra H_1 H_1 Z \ra b\bar b b\bar b \nu\bar \nu$ are searched for.
No indication of these processes is observed in the data as shown in Fig.~\ref{fig:cpvio}.
In general, CP-mixing reduces the MSSM mass limits significantly (center and right plots).
\vspace*{-0.4cm}
\clearpage

\vspace*{-0.4cm}
\section{Invisible Higgs Boson Decays}
\vspace*{-0.2cm}

\begin{figure}[htb]
\vspace*{-0.6cm}
\begin{center}
\includegraphics[scale=0.265]{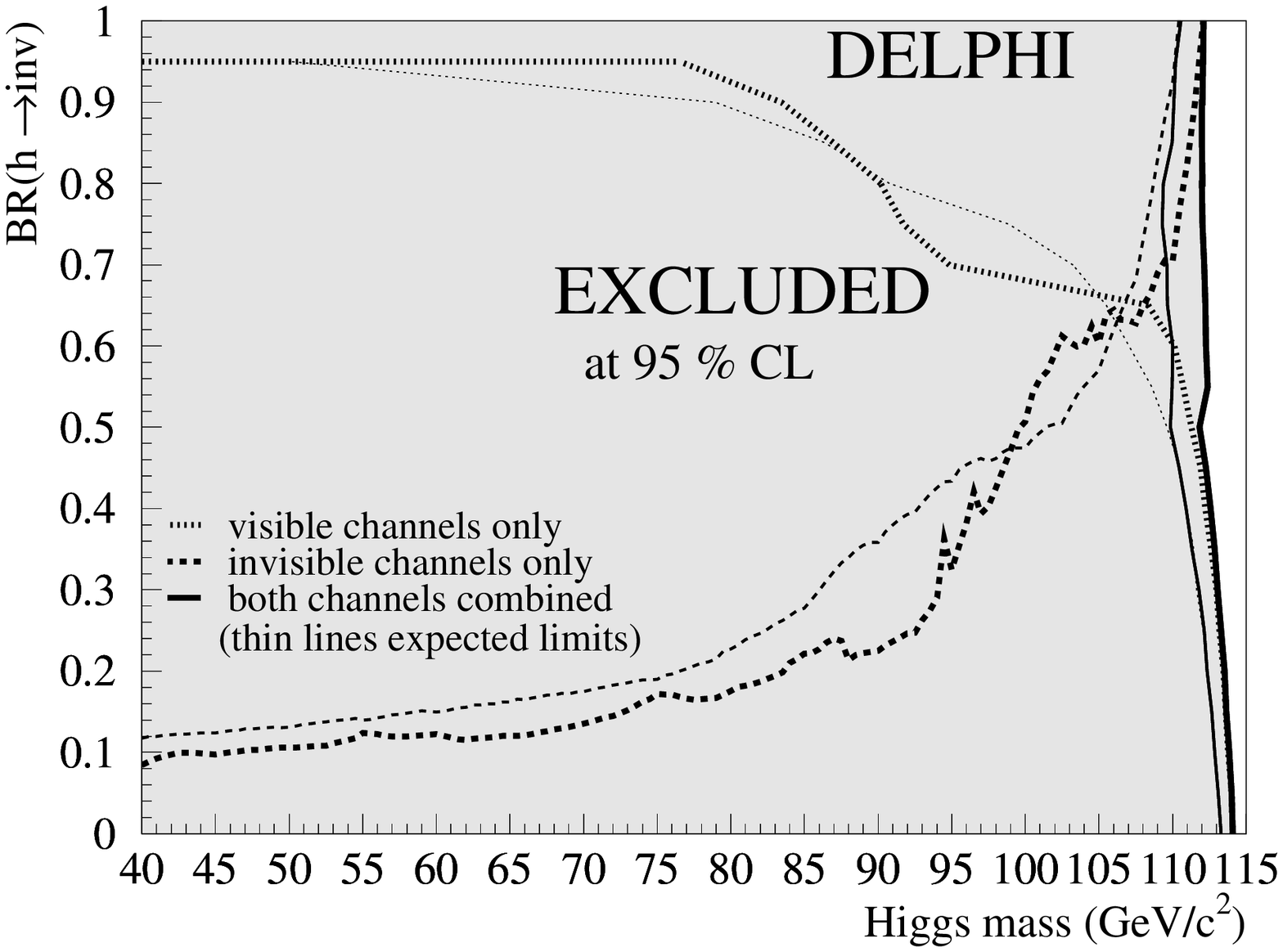}\hfill
\includegraphics[scale=0.195]{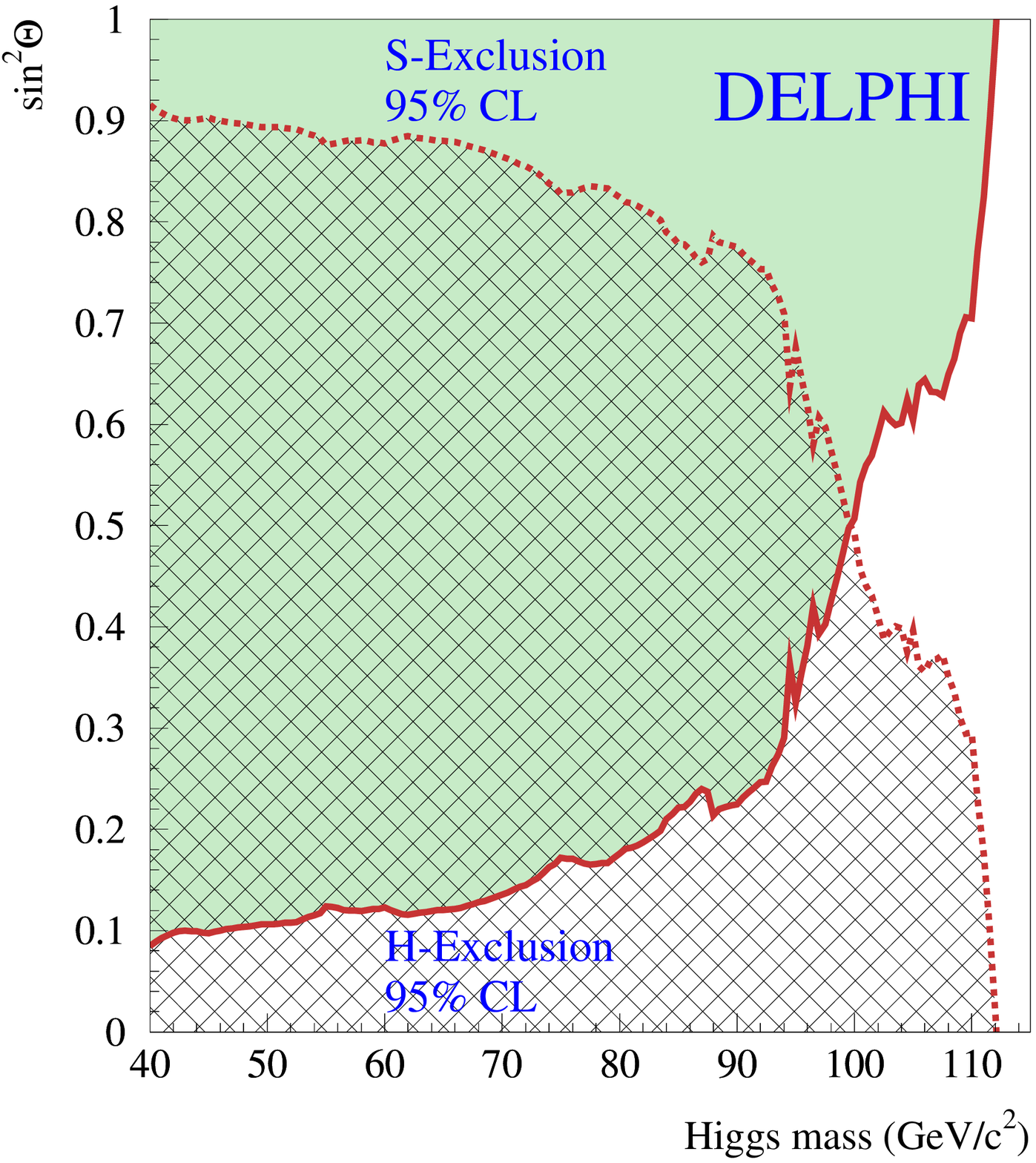} \hfill
\includegraphics[scale=0.195]{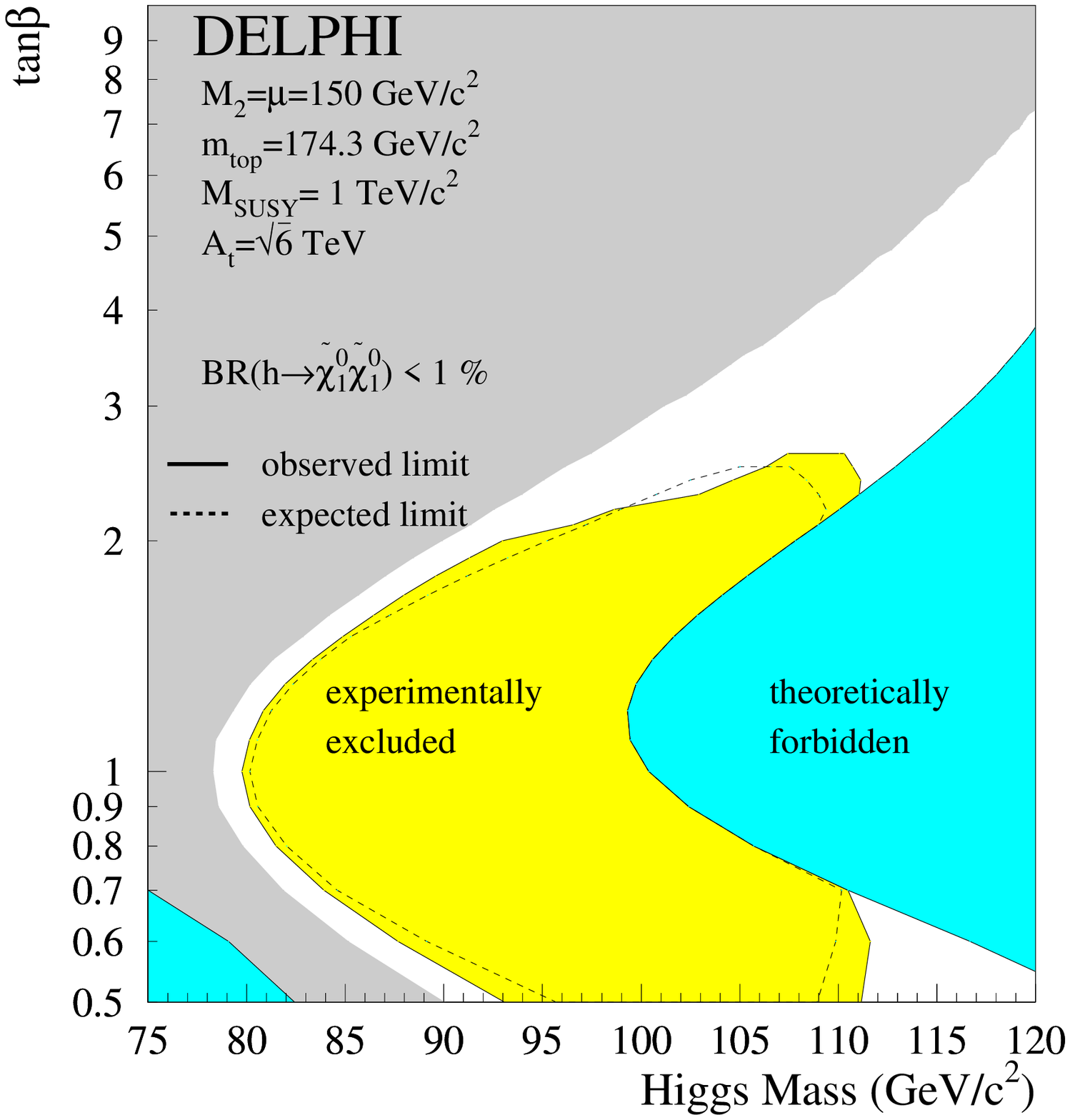}
\vspace*{-0.3cm}
\end{center}
\caption[]{Left: mass limits for SM and invisible Higgs boson decays combined.  
           Center: mass limits in Majoron models with an extra complex singlet,
           $\rm H/S\ra JJ$, where J escapes undetected. $\sin\theta$ is the H/S
           mixing angle.
           Right: mass limits in the MSSM for $\rm h\to \tilde\chi^0_1\tilde\chi^0_1$.
\label{fig:invlimit}}
\vspace*{-1.05cm}
\end{figure}

No indication of invisibly-decaying Higgs bosons is observed.
Figure~\ref{fig:invlimit} shows mass limits for SM and invisible Higgs boson decays combined.
The results are also interpreted in a 
Majoron model with an extra complex singlet, $\rm H/S\ra JJ$,
where J escapes undetected. In addition, mass limits are shown in the MSSM.

\section{Neutral Higgs Bosons in the General Two-Doublet Higgs Model}
\vspace*{-0.2cm}

\begin{figure}[htb]
\vspace*{-0.4cm}
\begin{center}
\includegraphics[scale=0.29]{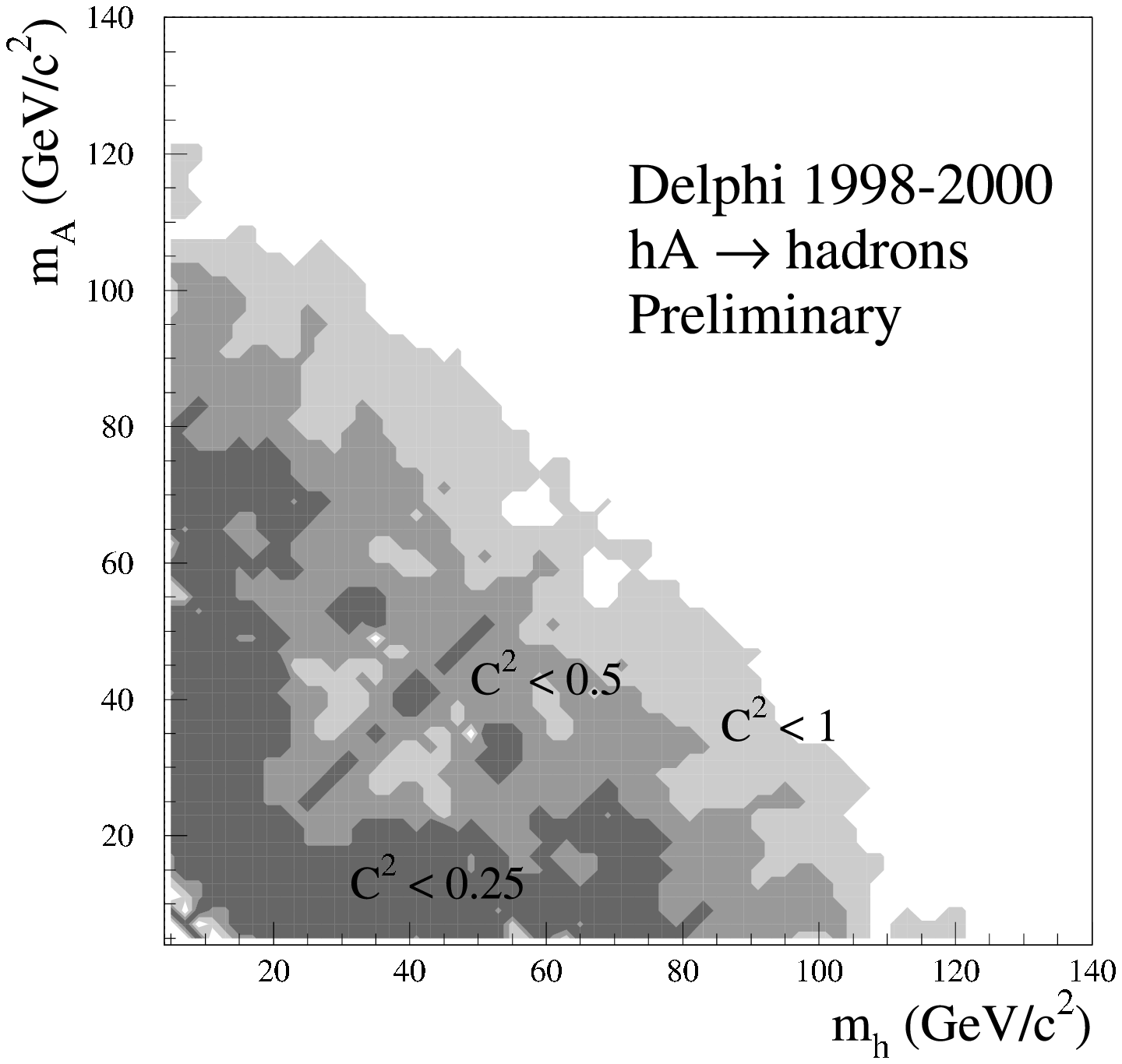}
\includegraphics[scale=0.195]{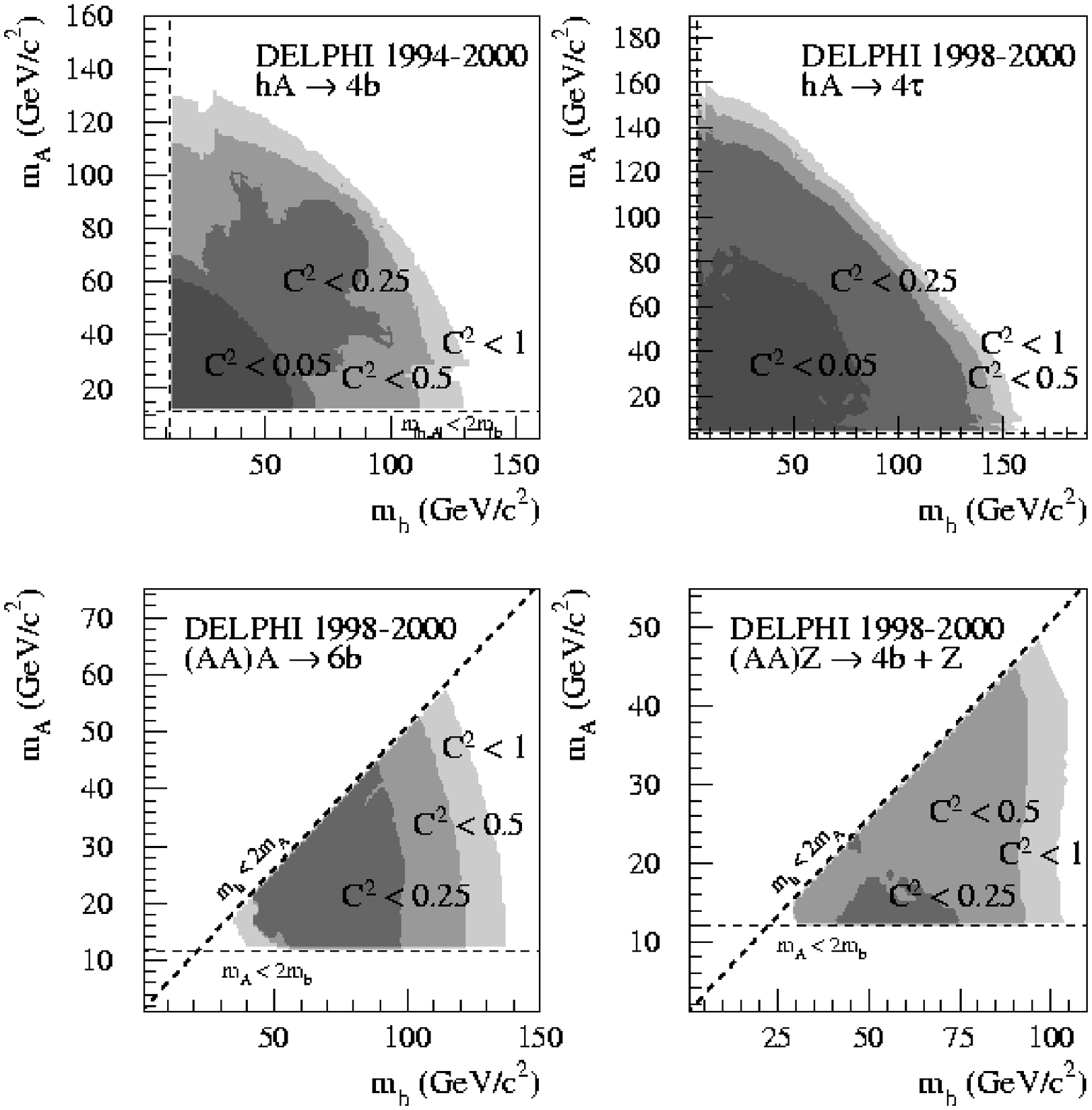}\hfill
\includegraphics[scale=0.22]{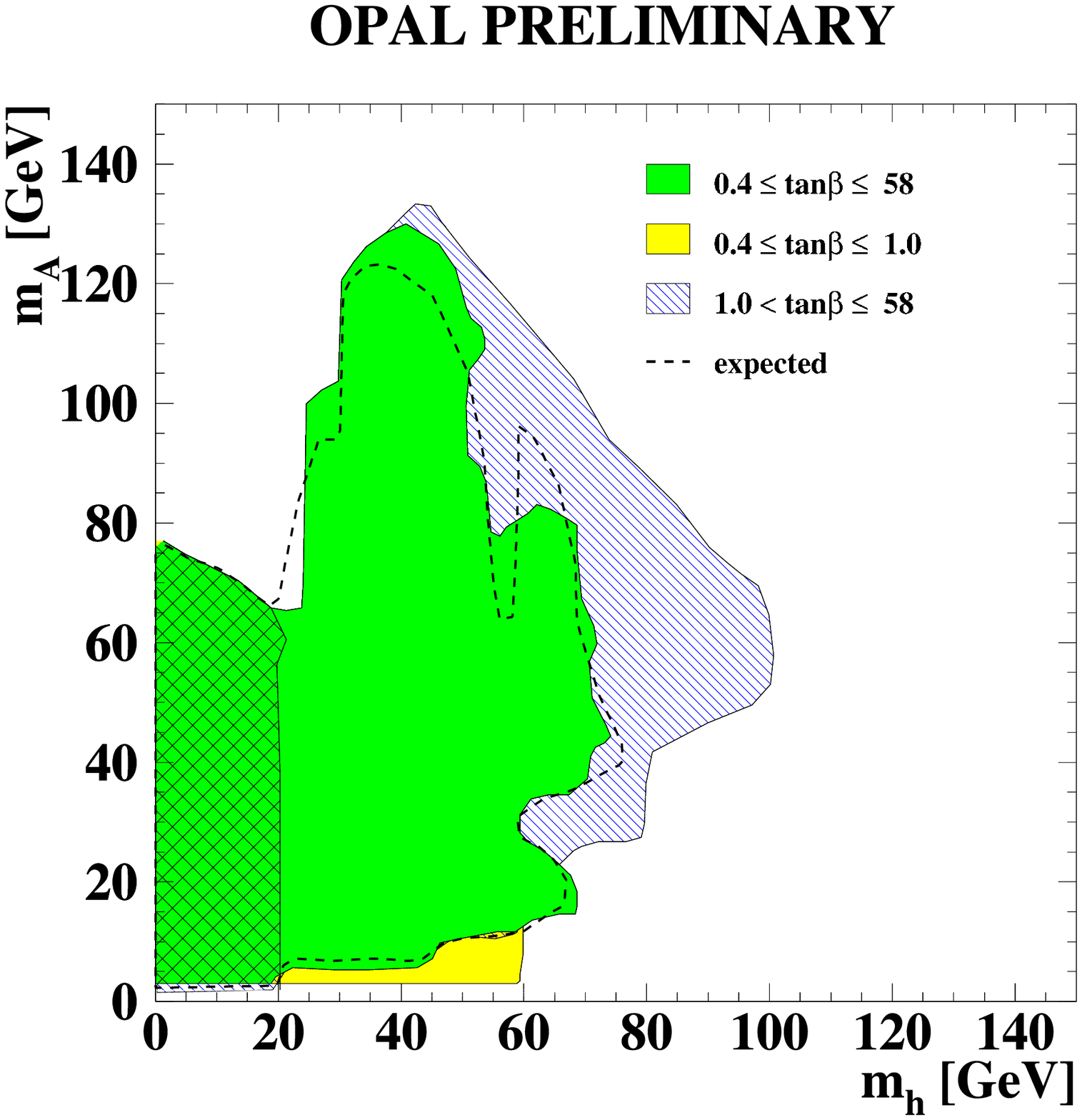}
\vspace*{-0.3cm}
\end{center}
\caption[]{Two-doublet Higgs model. 
           Left: Flavor-independent limits.            
           $C^2$ is the reduction factor on the maximum production cross section.
           Center:  Limits from dedicated searches for hA production.
           Right: mass limits from a general parameter scan in the two-doublet Higgs model.
\label{fig:2dhm}}
\vspace*{-0.8cm}
\end{figure}

Figure~\ref{fig:2dhm} shows mass limits from flavor-independent and
dedicated searches for hA production, and from a parameter scan.
The scan combines searches with b-tagging and flavor-independent searches.

\clearpage

\section{Yukawa Higgs Boson Processes 
\boldmath$\rm b\bar b h$ and $\rm b\bar b A$\unboldmath}
\vspace*{-0.2cm}

\begin{figure}[htb]
\vspace*{-0.4cm}
\begin{center}
\includegraphics[scale=0.43]{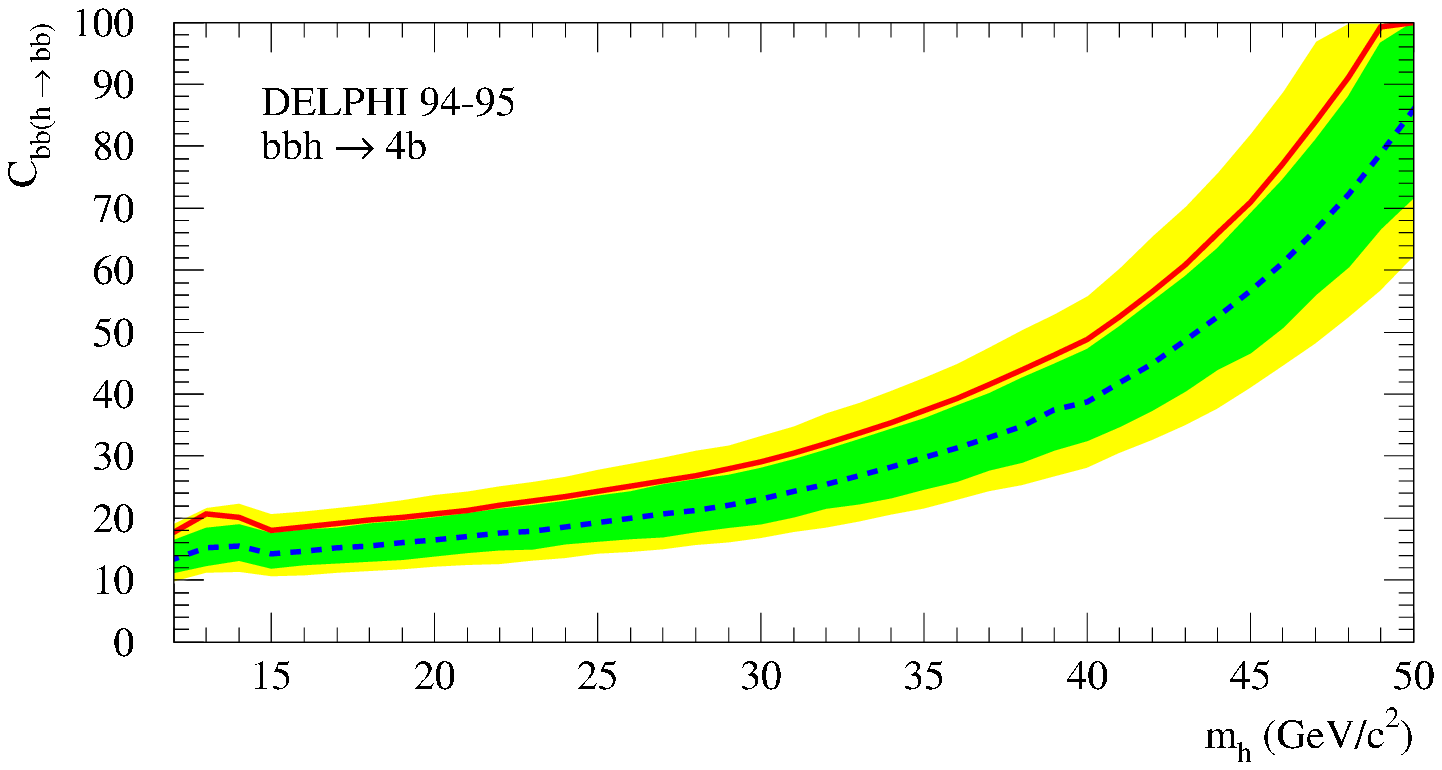}\hfill
\includegraphics[scale=0.43]{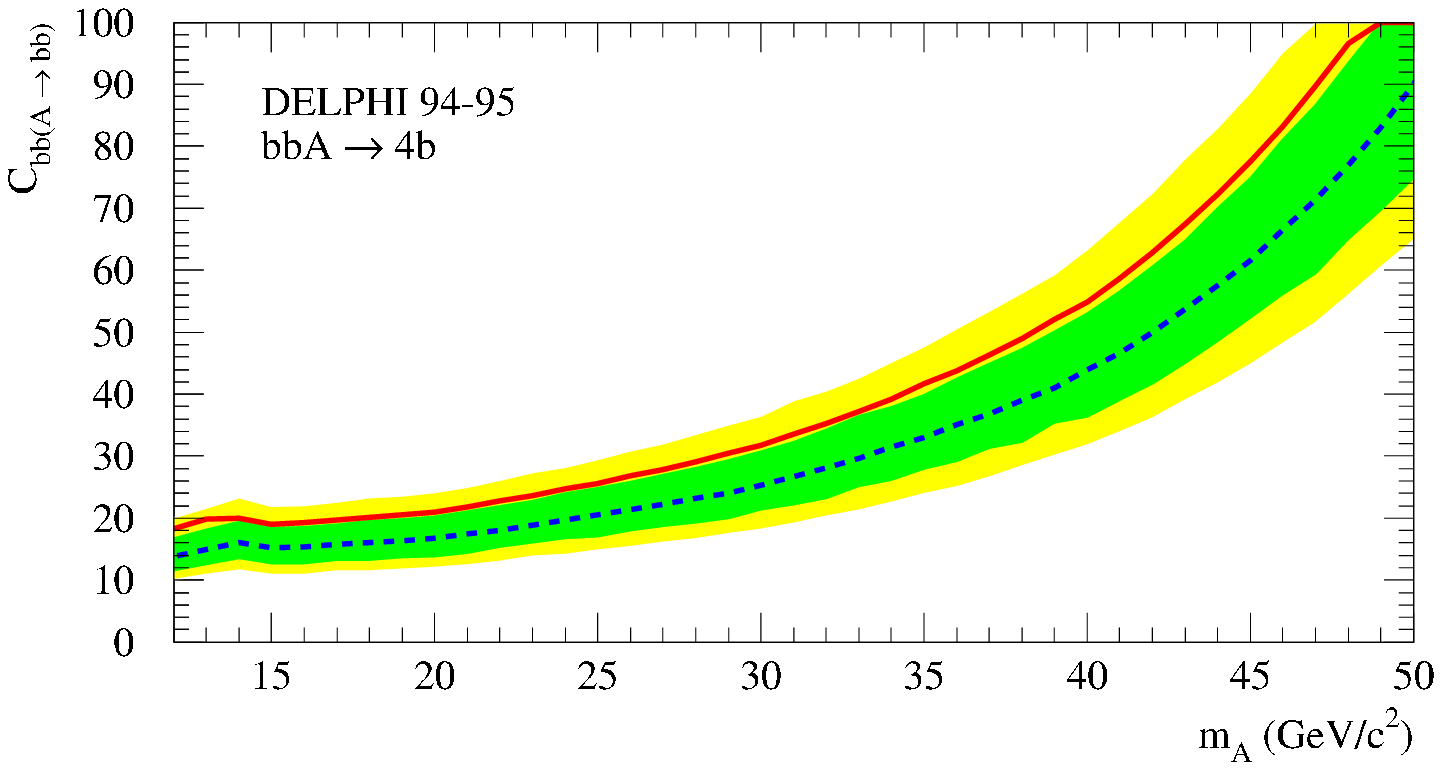}
\includegraphics[scale=0.43]{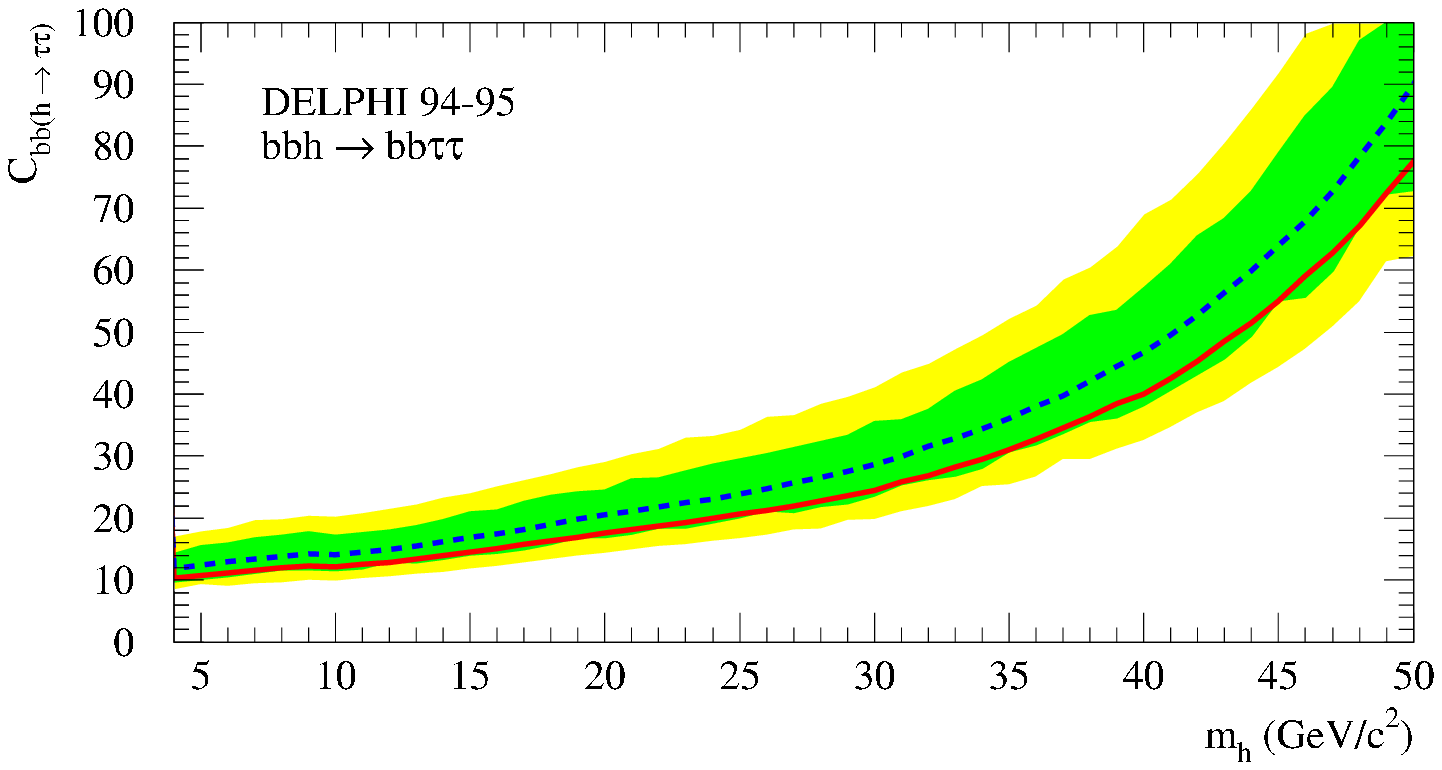}\hfill
\includegraphics[scale=0.43]{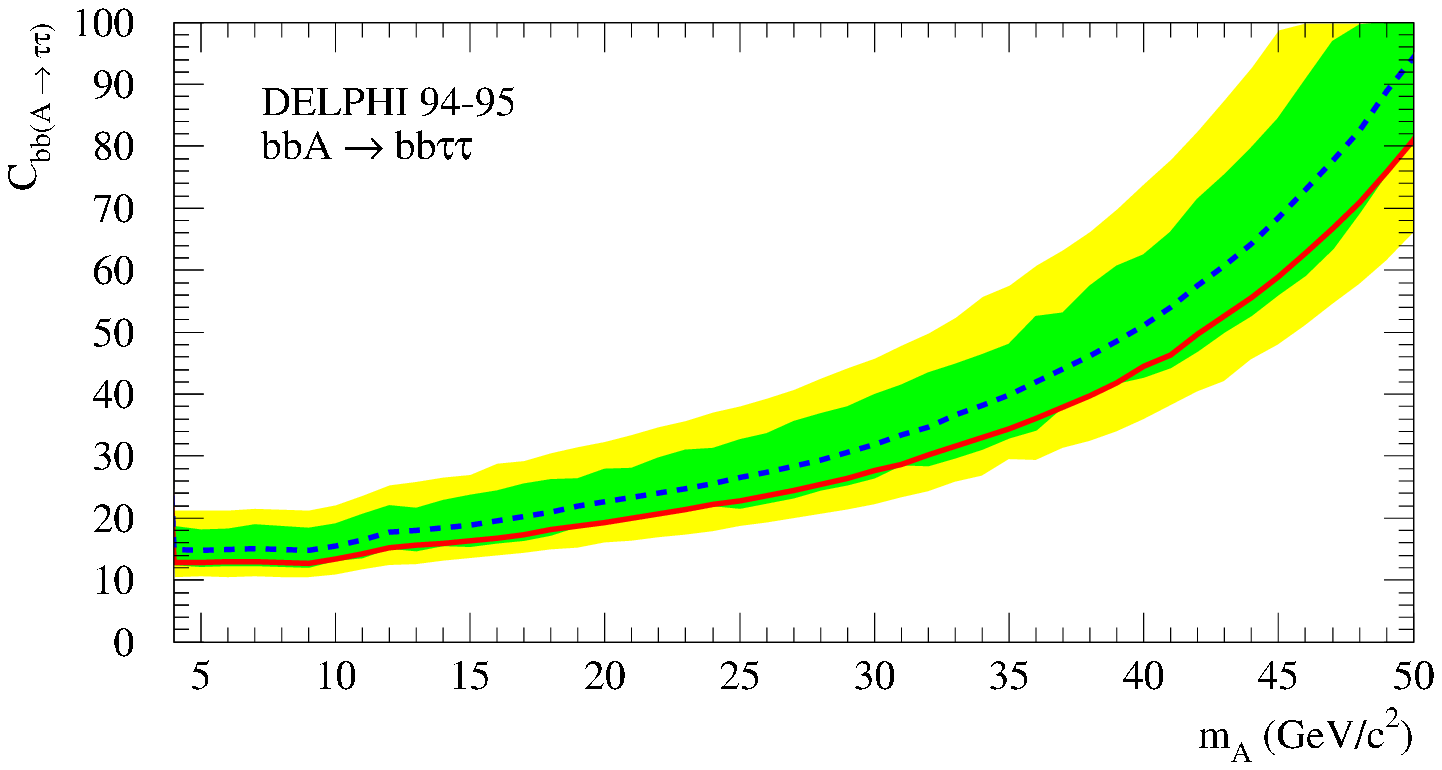}
\end{center}
\vspace*{-0.3cm}
\caption[]{Observed (solid line) and expected (dotted line) mass limits from searches 
           for the Yukawa processes 
           $\rm e^+e^-\rightarrow b\bar b\rightarrow b\bar bh,~b\bar bA$.
The $C$ factors include vertex enhancement factors and decay branching
fractions. The 1$\sigma$ and 2$\sigma$ error bands on the expected limit for background
are indicated (shaded areas).
\label{fig:yukawa}}
\vspace*{-0.5cm}
\end{figure}

Figure~\ref{fig:yukawa} shows mass limits from searches for the Yukawa processes 
$\rm e^+e^-\rightarrow b\bar b \rightarrow b\bar bh,~b\bar bA$.

\vspace*{-0.4cm}
\section{Singly-Charged Higgs Bosons}
\vspace*{-0.2cm}

\begin{figure}[htb]
\begin{center}
\vspace*{-0.8cm}
\includegraphics[scale=0.24]{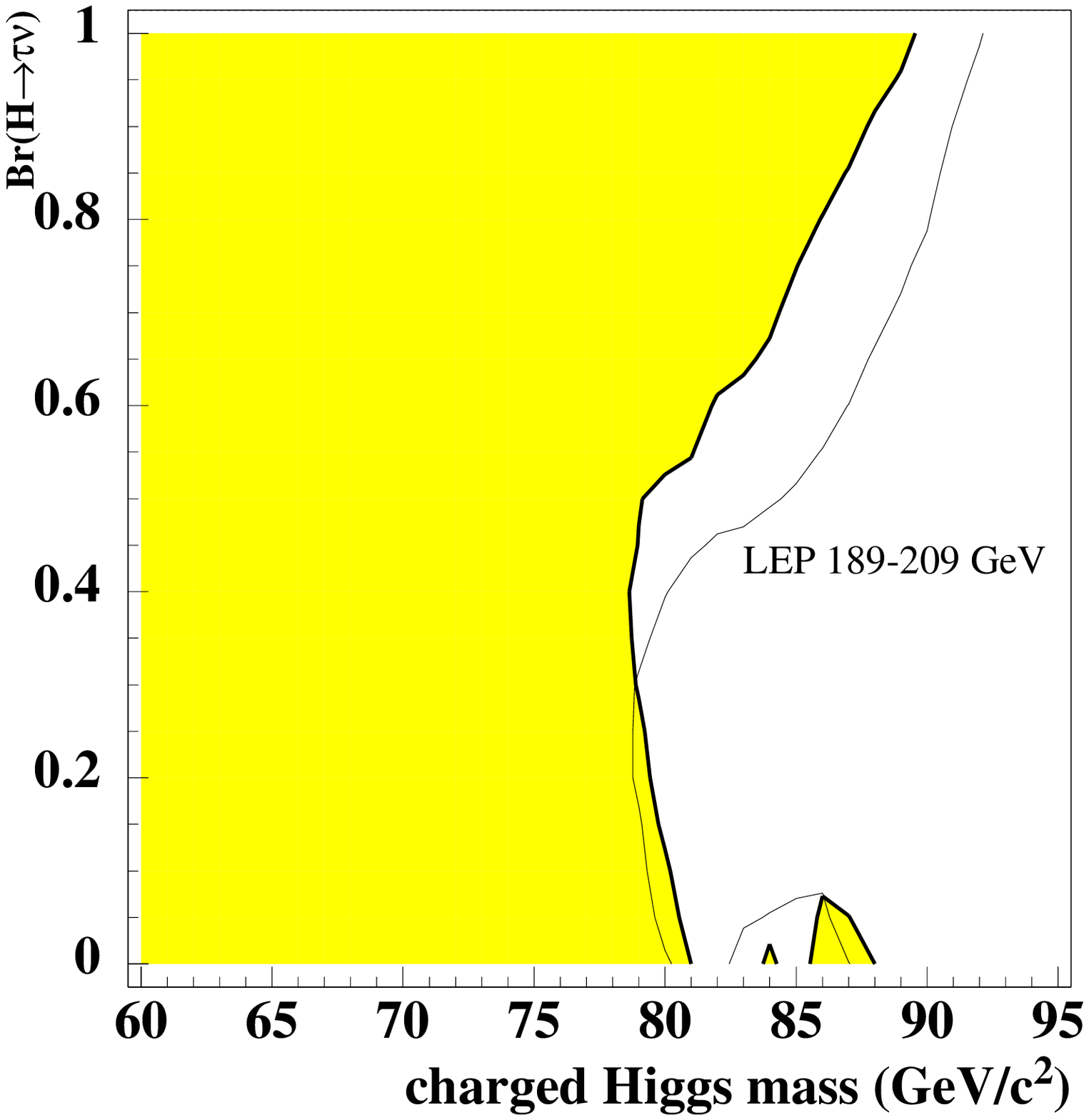}\hfill
\includegraphics[scale=0.22]{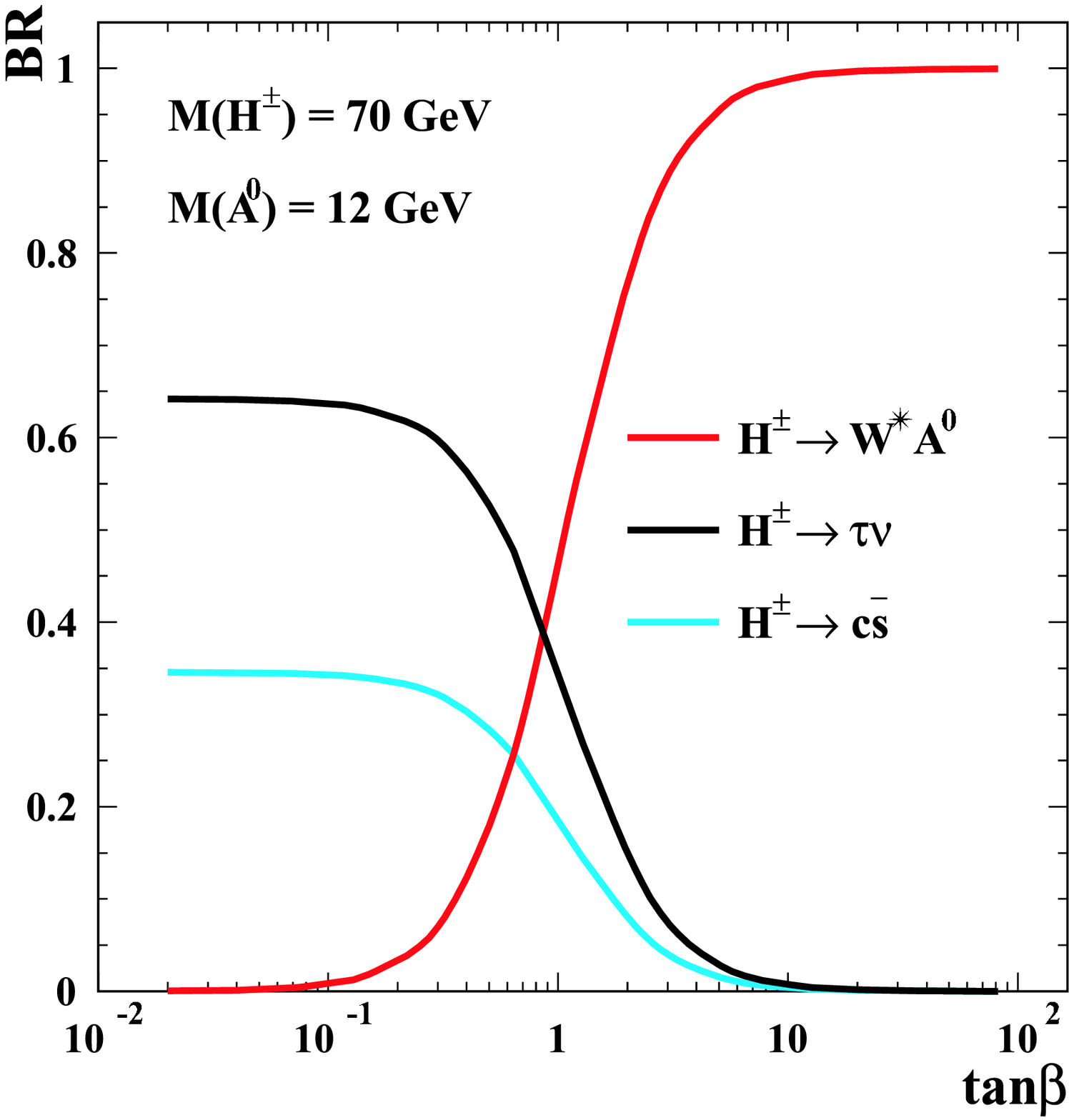}\hfill
\includegraphics[scale=0.22]{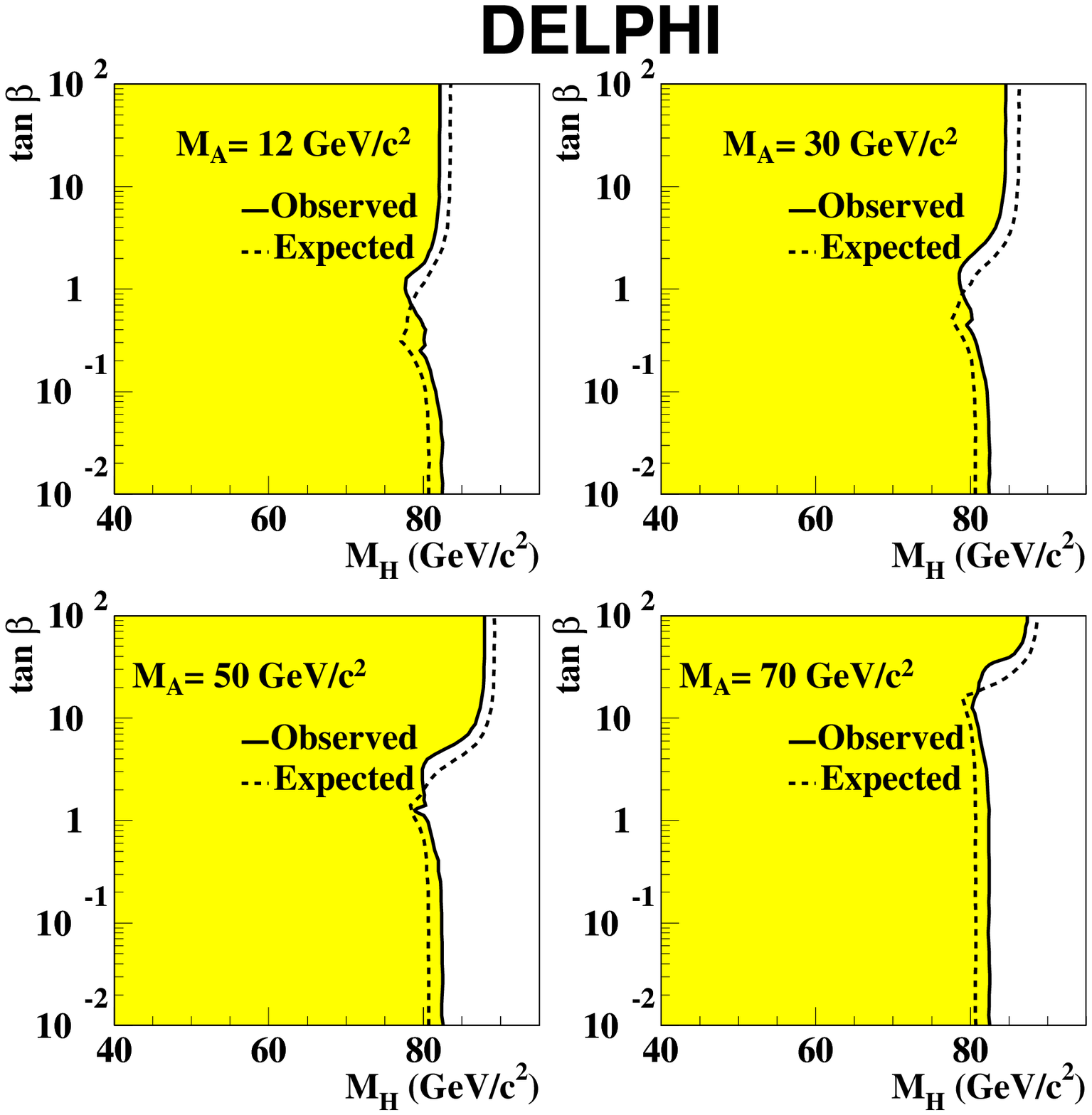}
\end{center}
\vspace*{-0.3cm}
\caption[]{Left: Excluded mass region (shaded area) from searches for
           $\rm \ee\ra H^+H^- \ra c\bar s\bar c s,$ $\rm cs\tau\nu$ and
           $\rm \tau^+\nu \tau^-\bar\nu$.
           The thin line shows the expected limit.
           Center: $\rm H^\pm \ra W^\pm A$ decays could be dominant for light A 
           boson masses.
           Right: excluded mass region (shaded area) from searches for this process.
\label{fig:hpm}}
\vspace*{-1.05cm}
\end{figure}

Figure~\ref{fig:hpm} shows mass limits from searches for
$\rm \ee\ra H^+H^- \ra c\bar s\bar c s,~cs\tau\nu,~\tau^+\nu \tau^-\bar\nu$.
The decay $\rm H^\pm \ra W^\pm A$ could be dominant and limits from 
dedicated\,searches\,are\,set.

\section{Doubly-Charged Higgs Bosons}
\vspace*{-0.2cm}

\begin{figure}[htb]
\vspace*{-0.45cm}
\begin{center}
\includegraphics[scale=0.23]{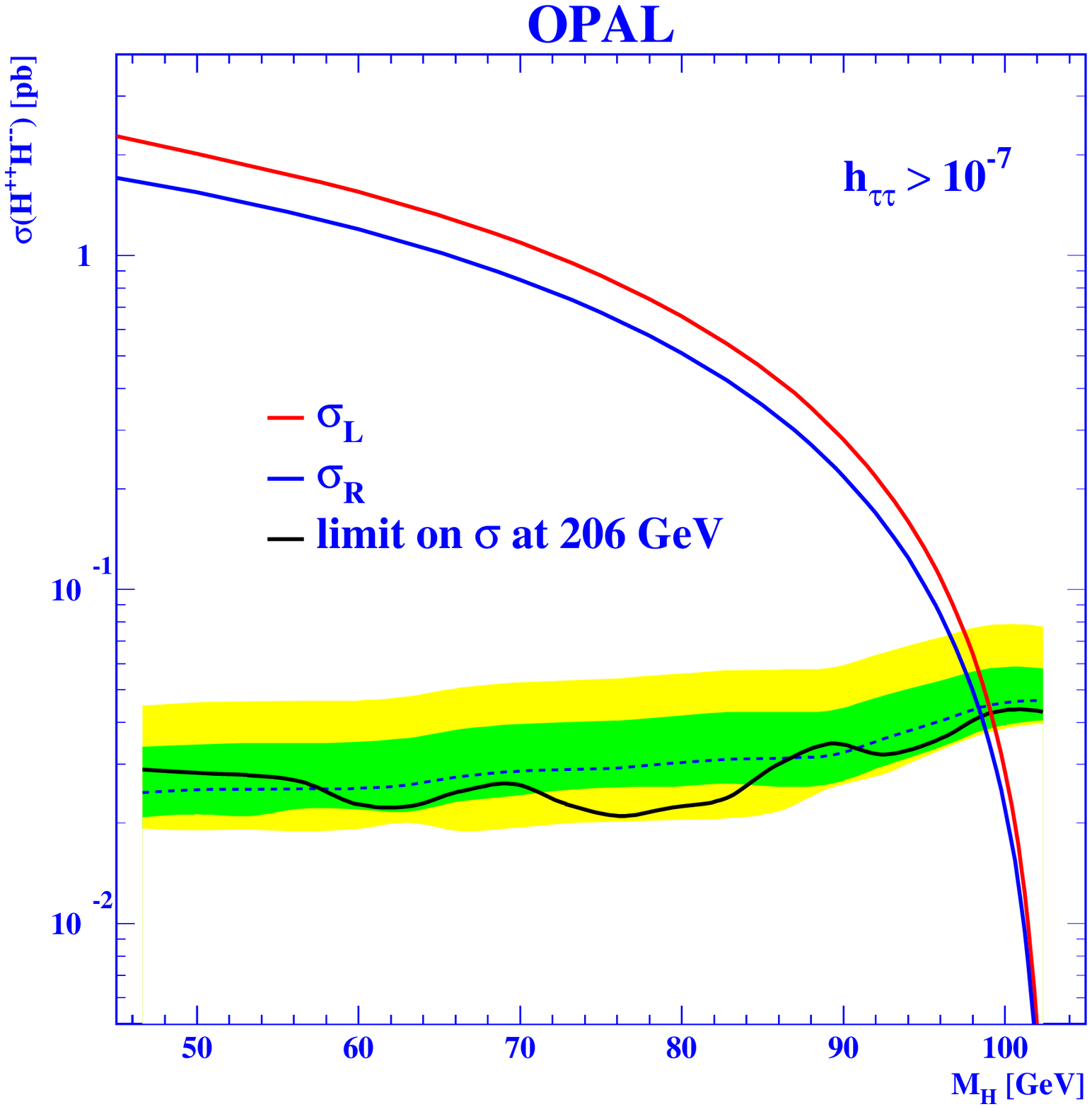}\hfill
\includegraphics[scale=0.30]{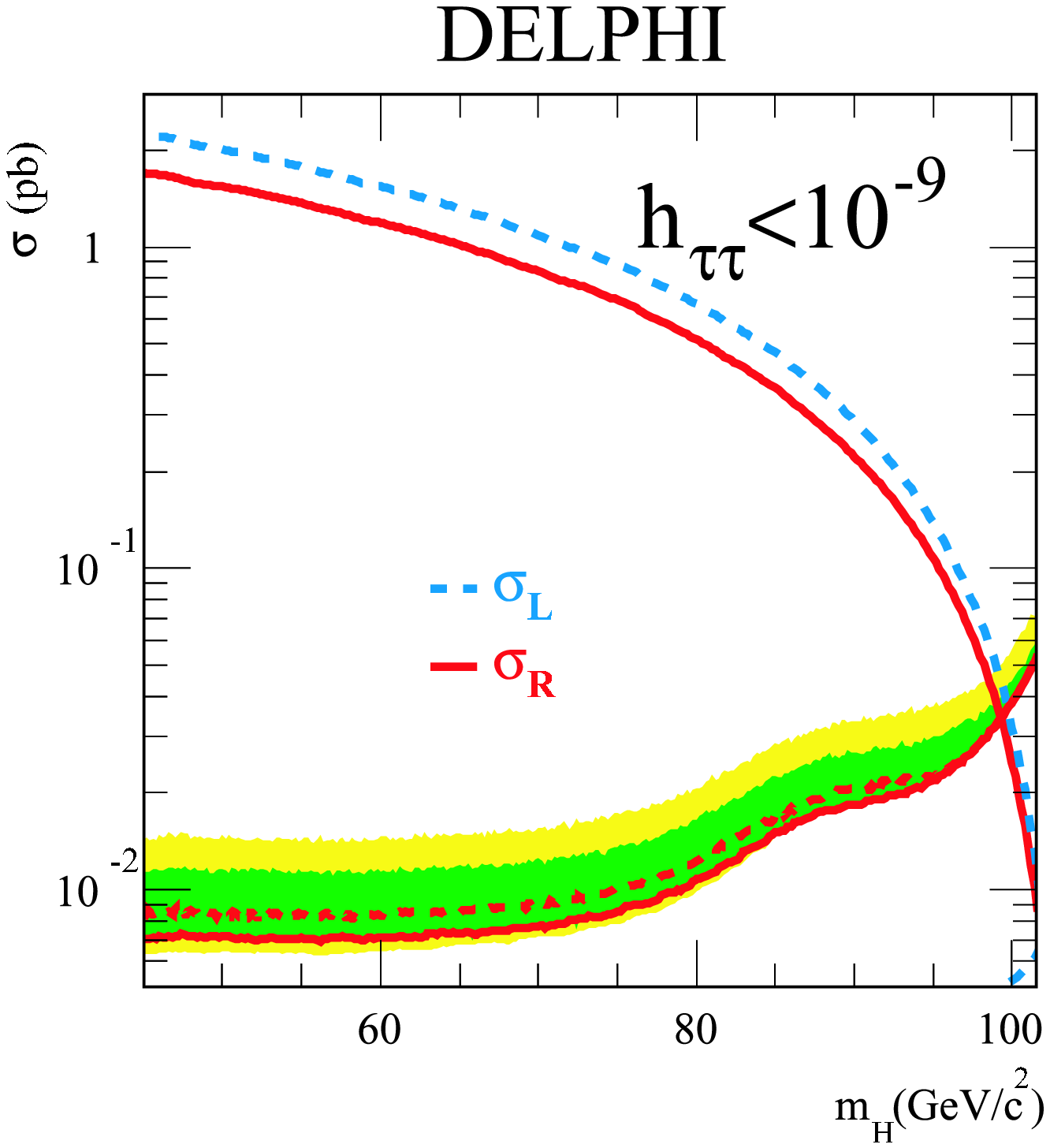}\hfill
\includegraphics[width=0.30\textwidth]{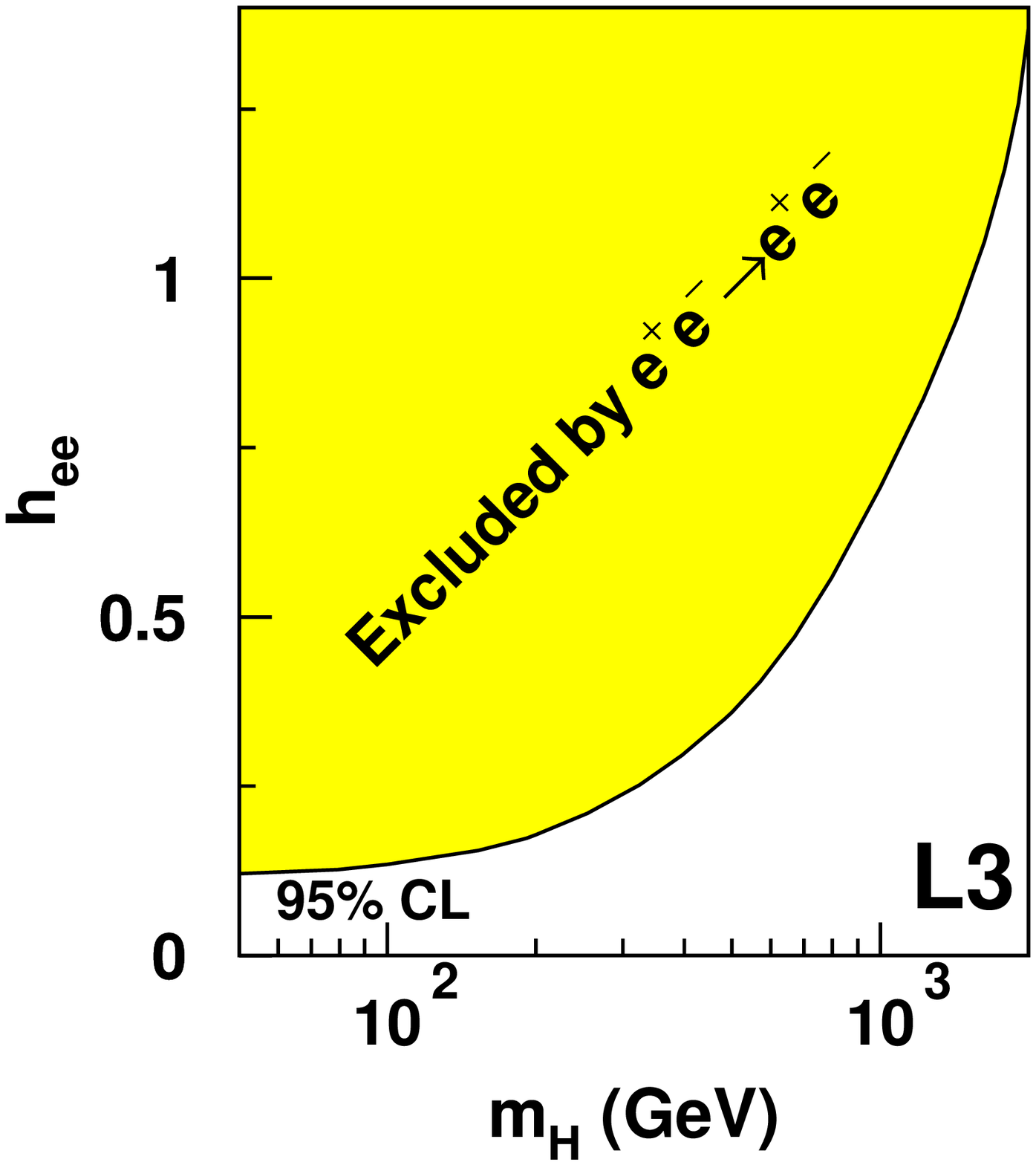}
\end{center}
\vspace*{-0.4cm}
\caption[]{Left and Center: Limits on the $\rm \ee\ra H^{++} H^{--}$ production cross section
           as a function of the doubly-charged Higgs boson mass. 
           The 1$\sigma$ and 2$\sigma$ error bands on the expected limit for background
           are indicated (shaded areas).
           Right: $\rm H^{--}$ limits from $\ee\ra\ee$ forward-backward asymmetry.
\label{fig:doubly}}
\vspace*{-0.5cm}
\end{figure}

The process $\rm \ee\ra H^{++} H^{--} \ra \tau^+\tau^+\tau^-\tau^-$ can lead
to decays at the primary interaction point 
($h_{\tau\tau}\geq 10^{-7}$) or a secondary vertex,
or to stable massive particle signatures.
Figure~\ref{fig:doubly} shows limits on the production cross section
and constraints by the forward-backward asymmetry of the process $\ee\ra\ee$.

\vspace*{-0.3cm}
\section{Fermiophobic Higgs Boson Decays:
\boldmath$\rm h\ra$ WW,~ZZ,~$\gamma\gamma$\unboldmath}
\vspace*{-0.2cm}

If Higgs boson decays into fermions are suppressed,
$\rm h\ra$ WW,~ZZ,~$\gamma\gamma$ decays could be
dominant. Mass limits from dedicated searches are set~\cite{dis_fermio}.

\vspace*{-0.2cm}
\section{Uniform and Stealthy Higgs Boson Scenarios}
\vspace*{-0.2cm}

\begin{figure}[htb]
\vspace*{-0.4cm}
\begin{minipage}{0.32\textwidth}
\includegraphics[width=1\textwidth]{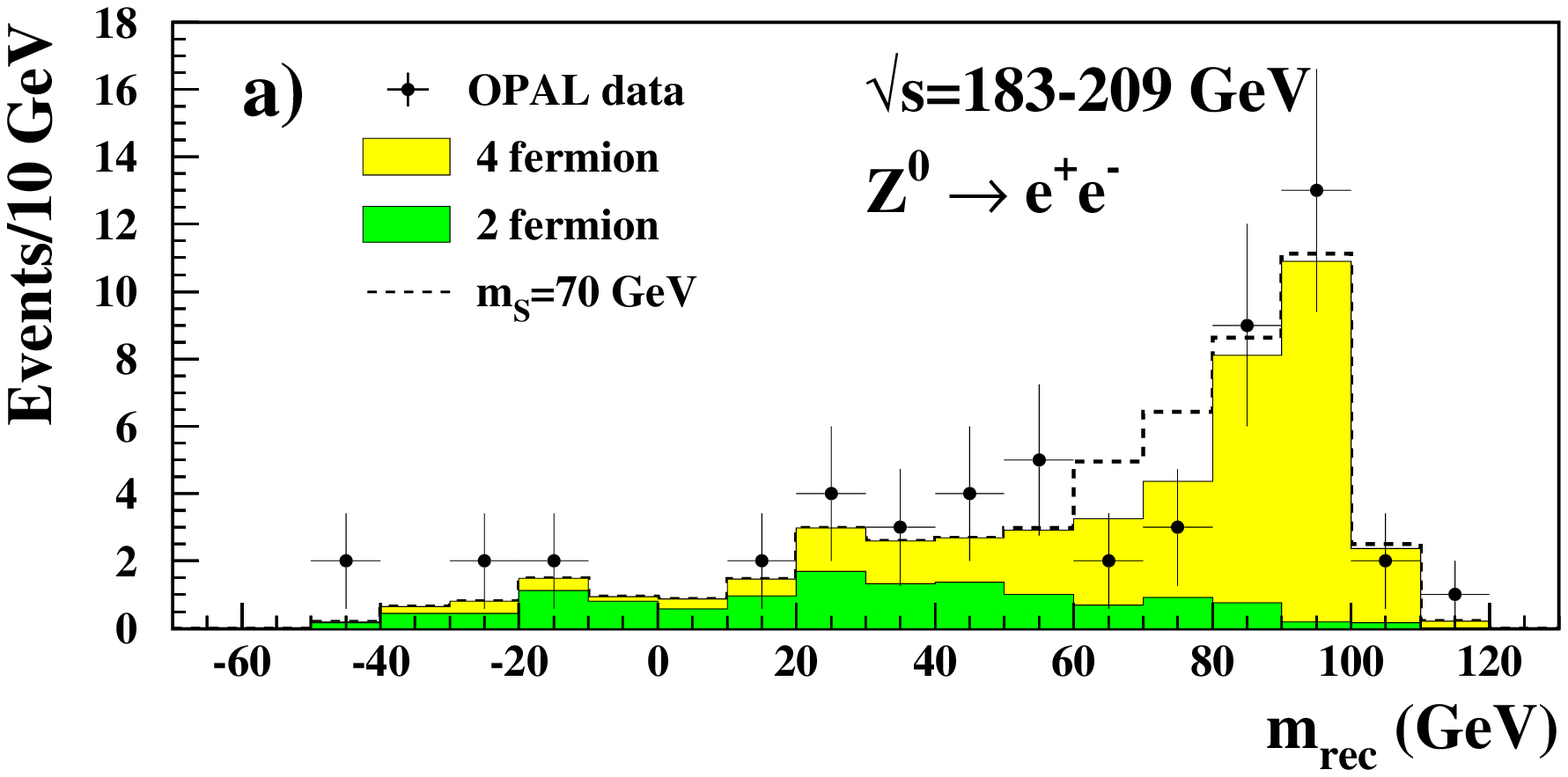}
\includegraphics[width=1\textwidth]{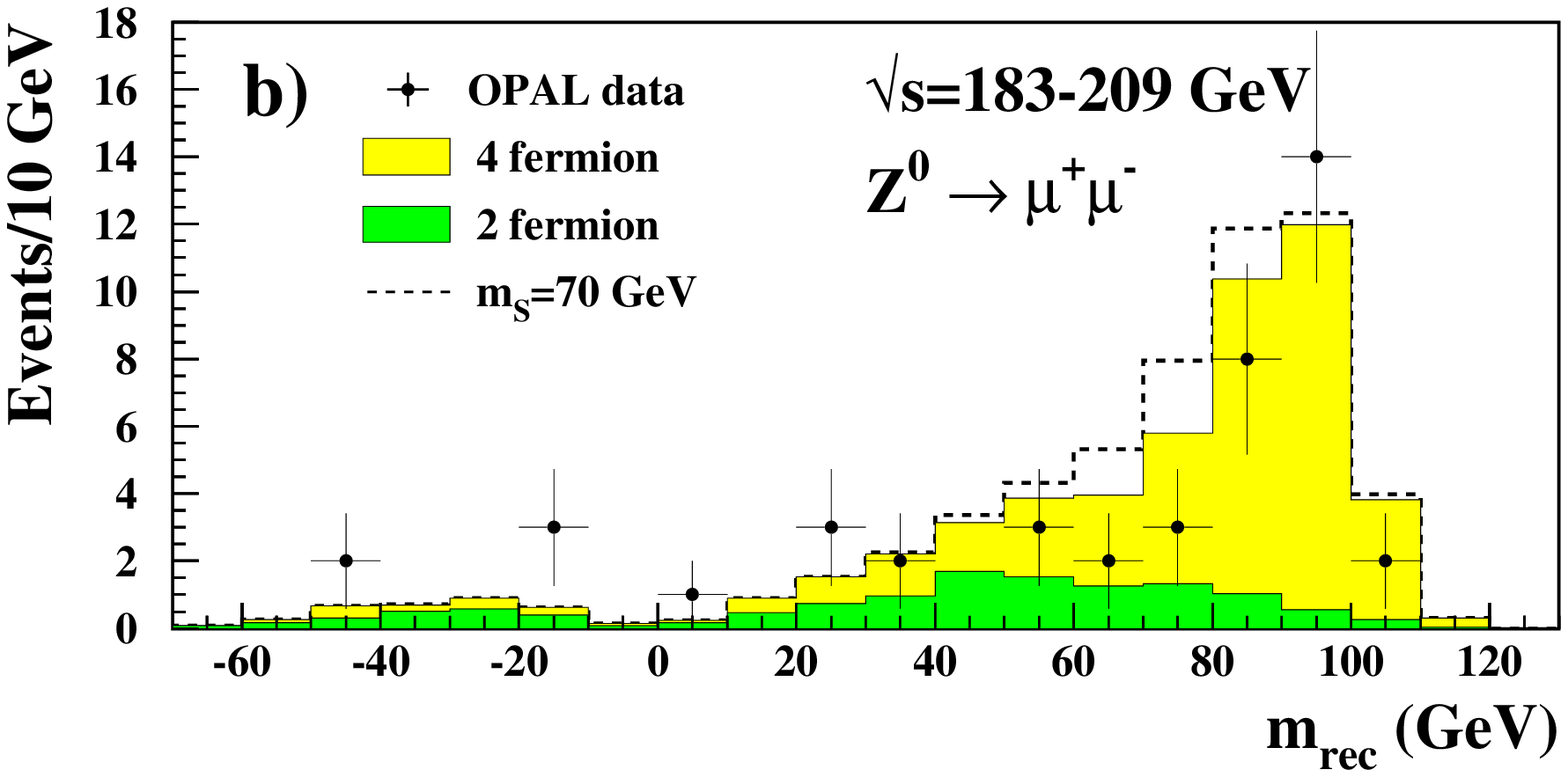}
\end{minipage}
\begin{minipage}{0.67\textwidth}
\includegraphics[width=0.50\textwidth]{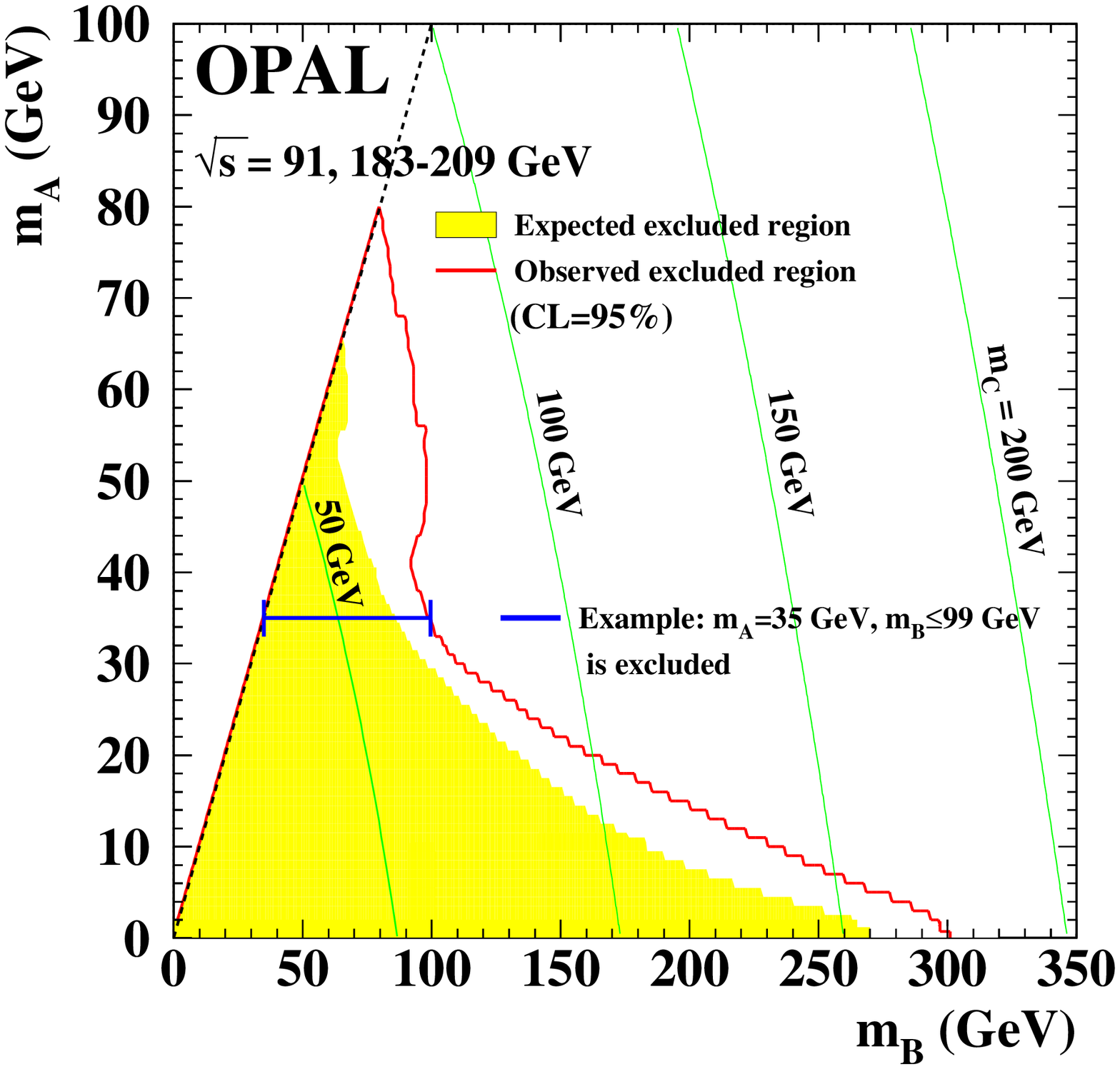}\hfill
\includegraphics[width=0.49\textwidth]{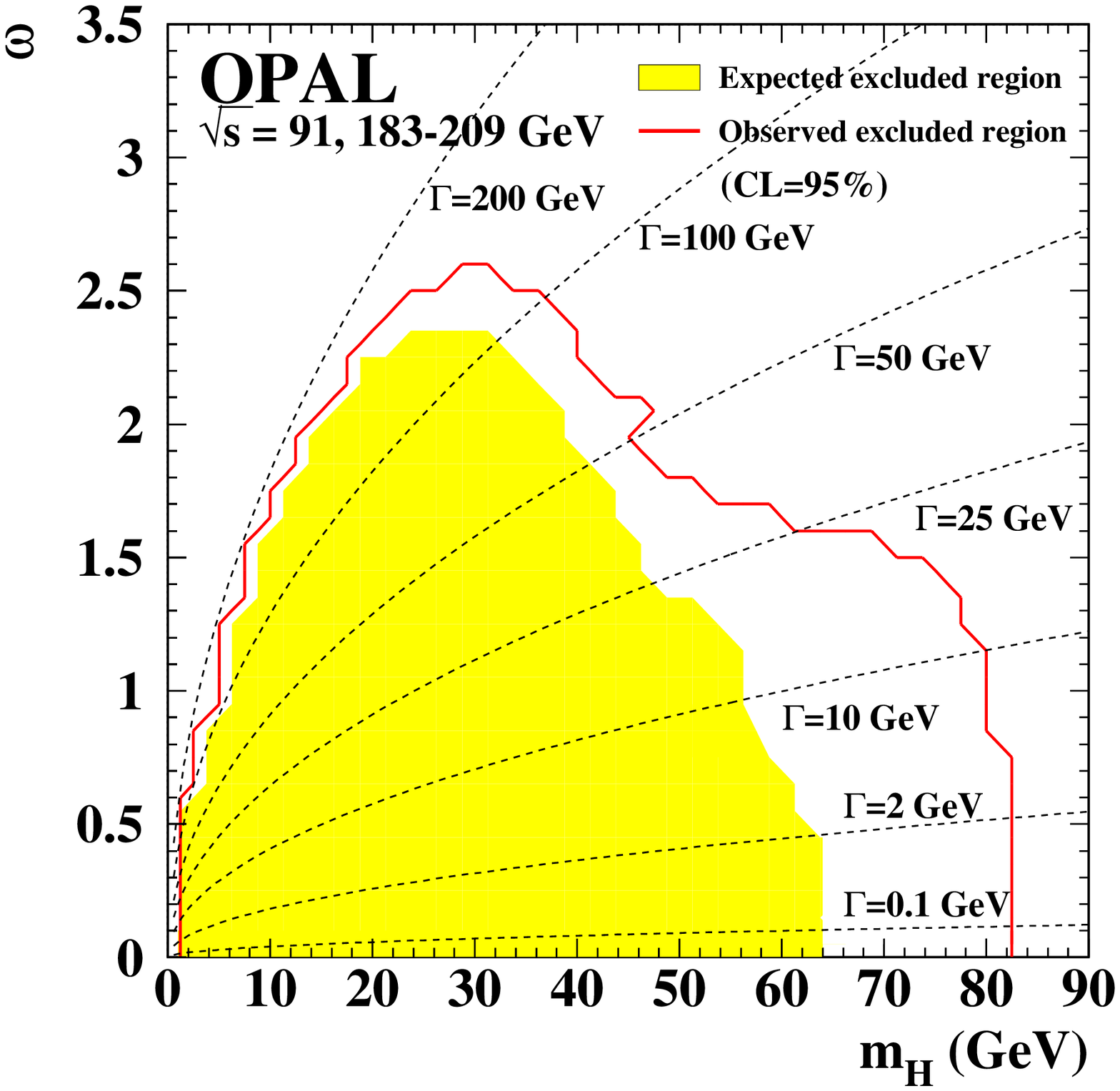}
\end{minipage}
\vspace*{-0.4cm}
\caption{Left: Recoiling mass spectrum  of $\rm Z\to e^+e^-$ and $\rm Z\to\mu^+\mu^-$.
         Center: Excluded mass range in the uniform Higgs model.
         Right: Mass limits in the stealthy Higgs model.
\label{fig:reco_ee}}
\vspace*{-0.7cm}
\end{figure}

The recoiling mass of the Z boson in the reaction $\rm e^+e^- \to HZ$
allows to search for the Higgs boson independent of the Higgs boson decay mode.
No indication of a Higgs boson signal has been observed as shown in Figs.~\ref{fig:reco_ee}.
Mass limits are shown in the uniform Higgs boson model,
where many uniform Higgs boson states exist in the range between
$m_{\rm A}$ and $m_{\rm B}$.
Another result from the recoiling mass spectrum is shown, where a stealthy Higgs boson 
has a large decay width owing to extra Higgs boson singlets in the model. The decay width 
depends on the parameter $\omega$.

\vspace*{-0.4cm}
\section{Conclusions }
\vspace*{-0.2cm}

Immense progress over a period of 15 years has been made in 
searches for Higgs bosons and much knowledge has been gained in preparation 
for new searches. 
No 
signal\,has\,been\,observed\,and\,various\,stringent\,limits\,are\,set\,as\,summarized\,in\,Table~\ref{tab:summary}. 

\vspace*{-0.3cm}
\section*{Acknowledgments}
\vspace*{-0.2cm}
I would like to thank the organizers of the DIS'2004\,conference\,for\,their\,kind\,hospi\-tality,
and colleagues from the LEP experiments for providing their latest results.

\vspace*{-0.5cm}
\mbox{ } 
\clearpage

\begin{table}[htb]
\small
\renewcommand{\arraystretch}{1.1} % enlarge line spacing
\vspace*{-0.6cm}
\caption{Summary of Higgs boson mass limits at 95\% CL.
`LEP' indicates a combination of the results from ALEPH, DELPHI, L3 and OPAL.
If results from the experiments are not (yet)\,com\-bined, examples
which represent the different research lines from individual experiments
are given.
\label{tab:summary} }
\vspace*{-0.7cm}
\begin{center}

\begin{tabular}{c|c|r}
 Search                     & Experiment & Limit \\\hline 
Standard Model              &   LEP  
   & $m^{\rm SM}_{\rm H} > 114.4$ GeV \\ 
Reduced rate and SM decay &       
  & $\xi^2>0.05:$ $ m_{\rm H} > 85$ GeV \\
& & $\xi^2>0.3:$ $ m_{\rm H} > 110$ GeV \\
Reduced rate and $\rm b\bar b$ decay  &   
  & $\xi^2>0.04:$ $ m_{\rm H} > 80$ GeV \\
& & $\xi^2>0.25:$ $ m_{\rm H} >110$ GeV \\ 
Reduced rate and $\tau^+\tau^-$ decay & 
  & $\xi^2>0.2:$ $ m_{\rm H} > 113$ GeV \\ 
\hspace*{-4mm} Reduced rate and hadronic decay\hspace*{-2mm} &
  & $\xi^2=1:$   $m_{\rm H} >112.9$ GeV\\ 
& & $\xi^2>0.3:$ $ m_{\rm H} > 97$ GeV \\ 
&ALEPH& $\xi^2>0.04:$ $m_{\rm H} \approx 90$ GeV \\ 
Anomalous couplings & L3 & $d,~\db,~\dgz,~\dkg$ exclusions \\ \hline
MSSM (no scalar top mixing) & LEP 
  & almost entirely excluded\\ 
General MSSM scan & DELPHI &  $m_{\rm h} > 87$ GeV, $m_{\rm A} >90$ GeV\\ 
Larger top-quark mass     & LEP & strongly reduced $\tan\beta$ limits \\ \hline
CP-violating models   & OPAL    &  strongly reduced mass limits  \\ \hline
Visible/invisible Higgs decays & DELPHI & $m_{\rm H} >111.8$ GeV\\ 
Majoron model (max. mixing) &  & $m_{\rm H,S} >112.1$ GeV\\ \hline
Two-doublet Higgs model   & DELPHI
  & $\rm hA\ra b\bar b b\bar b:$
    $m_{\rm h}+m_{\rm A} >  150$ GeV\\
(for $\sigma_{\rm max}$) & 
  & $\tau^+\tau^-\tau^+\tau^-:$
    $m_{\rm h}+m_{\rm A} >  160$ GeV\\
& & $\rm (AA)A\ra 6b:$ $m_{\rm h}+m_{\rm A} >  150$ GeV\\
& & $\rm (AA)Z\ra 4b~Z:$ $m_{\rm h} >  90$ GeV\\
& & $\rm hA\ra q\bar q q\bar q:$ 
      $m_{\rm h}+m_{\rm A} >  110$ GeV\\
Two-doublet model scan & OPAL
  & $\tan\beta > 1:$ $ m_{\rm h} \approx m_{\rm A} > 85$ GeV \\\hline 
Yukawa process & DELPHI & $C > 40:$ $m_{\rm h,A} > 40$ GeV \\\hline 
Singly-charged Higgs bosons & LEP 
  & $m_{\rm H^\pm} > 78.6$ GeV \\
$\rm W^\pm A$ decay mode & DELPHI& $m_{\rm H^\pm} > 76.7$ GeV \\ \hline
Doubly-charged Higgs bosons & \hspace*{-2.5mm} DELPHI/OPAL \hspace*{-2.5mm} 
  & 
$m_{\rm H^{++}} > 99$ GeV \\
$\ee\ra\ee$ &L3 &$h_{\rm ee} > 0.5:$ $m_{\rm H^{++}} > 700$ GeV \\ \hline
Fermiophobic $\rm H\ra WW, ZZ, \gamma\gamma$ & L3 
  &  $m_{\rm H} > 108.3$ GeV \\
$\rm H\ra \gamma\gamma$ &LEP &  $ m_{\rm H} > 109.7$ GeV \\ \hline
Uniform and stealthy scenarios & OPAL & depending on model parameters
\end{tabular}
\end{center}
\vspace*{-0.6cm}
\end{table}


\vspace*{-0.2cm}
\begin{thebibliography}{0}
\vspace*{-0.2cm}

\bibitem{sm}
ALEPH, DELPHI, L3 and OPAL Collaborations
and the LEP working group for Higgs boson searches, 
Phys. Lett. {\bf B  565} (2003) 61.

\bibitem{summer2003}
ALEPH, DELPHI, L3 and OPAL Collaborations, contributed papers to the 
International Europhysics Conference on High Energy Physics
EPS, 17-23 July 2003, Aachen, Germany; and
XXI International Symposium on Lepton and Photon Interactions at High Energies,
11-16 August 2003, Fermi National Accelerator Laboratory, Batavia, Illinois, USA;
and recent updates.

\bibitem{dis04hepph}
A.~Sopczak, hep-ph/0408047.

\bibitem{as2000}
A.~Sopczak, Proc. DPF-2000, hep-ph/0011285;
            Proc. Topical seminar on the legacy of LEP and SLC, Siena,
            Nucl. Phys. Proc. Supp. {\bf 109} (2002) 271.

\bibitem{dis_fermio}
Ch.~Delaere, ``Fermiophobic Higgs'', these proceedings.

\vspace*{-0.4cm}
\end{thebibliography}
\end{document}